\documentclass[%
 aps,pre,
 amsmath,amssymb,superscriptaddress,
 reprint,%
]{revtex4-2}

\usepackage{graphicx}
\usepackage{dcolumn}
\usepackage{bm}
\usepackage{amsfonts,amsthm,mathbbol}
\usepackage[utf8]{inputenc}
\usepackage[T1]{fontenc}
\usepackage{mathptmx}

\usepackage{color}

\begin{document}

\title[Thermal effects on fracture and brittle-to-ductile transition]{Thermal effects on fracture and brittle-to-ductile transition}

\author{Andrea Cannizzo}%
 \email{andrea.cannizzo@iemn.fr}
 \affiliation{Univ. Lille, CNRS, Centrale Lille, Univ. Polytechnique Hauts-de-France, UMR 8520 - IEMN - Institut d’Électronique de Microélectronique et de Nanotechnologie, F-59000 Lille, France}
\affiliation{Politecnico di Bari, (DMMM) Dipartimento di Meccanica, Matematica e Management, Via Re David 200, I-70125 Bari, Italy}

\author{Stefano Giordano}%
 \email{stefano.giordano@univ-lille.fr}
    \affiliation{Univ. Lille, CNRS, Centrale Lille, Univ. Polytechnique Hauts-de-France, UMR 8520 - IEMN - Institut d’Électronique de Microélectronique et de Nanotechnologie, F-59000 Lille, France}

\date{\today}

\begin{abstract}
The fracture behavior of brittle and ductile materials can be strongly influenced by thermal fluctuations, especially in micro- and nano-devices as well as in rubberlike and biological materials. However, temperature effects, in particular on the brittle-to-ductile transition,  still require a deeper theoretical investigation. As a step in this direction we propose a theory, based on equilibrium statistical mechanics, able to describe the temperature dependent brittle fracture and brittle-to-ductile transition in prototypical discrete systems consisting in a lattice with breakable  elements. Concerning the brittle behavior, we obtain closed form expressions for the temperature-dependent fracture stress and strain, representing a generalized Griffith criterion, ultimately describing the fracture as a genuine phase transition. With regard to the brittle-to-ductile transition, we obtain a complex critical {\it scenario} characterized by a threshold temperature between the two fracture regimes (brittle and ductile), an upper and a lower yield strength, and a critical temperature corresponding to the complete breakdown.
To show the effectiveness of the proposed models in describing thermal fracture behaviors at small scales,
we successfully compare our theoretical results with molecular dynamics simulations of Si and GaN nanowires.

\end{abstract}

\maketitle

\section{Introduction}

The mechanical degradation of a material typically results from the insurgence of cracks, from their geometric arrangement and interactions and, finally, from temperature.
The classical Griffith energetic approach in fracture mechanics deduces that, under a homogeneous stress $\sigma$, a single slit crack with half-length $L$ grows if $\sigma>\sqrt{2\gamma_s E'/(\pi L)}$,
while if $\sigma<\sqrt{2\gamma_s E'/(\pi L)}$, it remains stable \cite{griffith,pugno}. 
Here, $E'$ is the equivalent elastic modulus, equal to the Young modulus $E$ in plane stress condition, and equal to $E/(1-\nu^2)$ in plane strain condition, where $\nu$ is the Poisson ratio of the material. 
Moreover, $\gamma_s$ is the surface energy density, that is, the energy expended to debond a unit length crack. 
This stability criterion measures an energy competition between the free surface energy created by the fracture and the elastic energy stored in the deformable solid. 
The latter has been evaluated within the linear elasticity theory by Inglis \cite{Inglis}
and Kolosoff \cite{kolosoff} and used by Griffith to develop his criterion.
The ingenious approach proposed by Griffith  in his celebrated criterion has been largely  and successfully tested in glass and other brittle materials containing cracks of controlled length \cite{lawn,och}, and also validated by atomistic simulations in ideal mono-crystalline systems \cite{mattoni,china}.
Its main limiting hypothesis is that the overall fracture energy coincides with the surface energy, i.e. with the energy needed to break the bonds between the two crack faces.
Since Griffith's theory fails to apply to ductile materials (where fracture energy is much higher than the only surface energy  \cite{orowan1,orowan2}) it was generalized by Irwin to include plastic dissipation \cite{Irwin1,Irwin2}.
More advanced models for ductile fracture, taking into account an explicit description of the `cohesive zone' where the plastic processes localize, have then been proposed by Dugdale and Barenblatt \cite{dugdale,barenblatt}. 
The theoretical relationship between the Griffith's theory, its Irwin modification, and the Dugdale-Barenblatt models was initially studied by Willis \cite{willis} and further investigated by Rice through the concept of $J$-integral \cite{rice}. 
In addition to the single fracture study, an important topic in linear elastic fracture mechanics is represented by the collective degradation mechanism induced by populations of cracks that interact depending on their geometric arrangement \cite{kachanov1,kachanov2,giordaPRL,giordaEFM,giordaPRB,giordaPRB2,dormieux,giordaRCC,mancaPRL,mancaEPJE,dormieuxkondo}.
Due to the wide scientific and technological interest, the theoretical and experimental studies of fracture phenomena have been extremely extensive. Therefore, its history is long and complicated and, here, we refer the reader to the relevant literature \cite{broberg,ivanova,brobergbook,gross,erdogan}.
We simply mention that current advanced researches concern the traction-separation relation in cohesive models \cite{park,ferdousi},  the instability in dynamic fracture \cite{marder1,chen}, and the variational approach to fracture \cite{marigo,khonsari}. Computational techniques for cracks propagation include the phase field method and the dual-horizon peridynamics formulation \cite{nume1,nume2,nume3,nume4}.

Previously described investigations are predominantly based on deterministic assumptions and theories.
Of course, also statistical approaches have been widely applied to rupture phenomena \cite{truski,zapperi,moreno,alava,kawamura,hansen}, and among others those based on the so-called fiber bundle model are particularly significant \cite{sornette,phoenix,pradhan,kun}.
Importantly, the statistical analysis plays a crucial role for understanding the effect of disorder in failure processes \cite{diso,zanzo,parisi,ponson,niccolini,sinha,kondo,rocha}.
Despite the wide diffusion of statistical techniques, the approaches that allow to study the effects of thermal fluctuations on the fracture are rather limited \cite{ausloos,khtrin,marder2,ortiz,ciliberto1,ciliberto2,ciliberto3,bonamy,cochard}.
In particular, the temperature dependence of crack stability criteria has not been studied explicitly.
For this reason, we propose here two paradigmatic models able to evaluate the effects of thermal fluctuations on the quasi-static brittle fracture and on the brittle-to-ductile transition.
These approaches make it possible to study how fracture stability is influenced by temperature changes and to determine the transition temperature between brittle and ductile behaviors.
The proposed fracture models are based on equilibrium statistical mechanics and they are implemented by means of the spin variable approach, useful to deal with arbitrarily non-convex potential energies \cite{soft}. 
This method has been largely applied to several situations including the physics of muscles \cite{car1,car2}, the folding of macromolecules \cite{jcpbenedito,beneditopre,florio,puglisi,romain}, the adhesion or peeling processes  \cite{prr,jpa}, the phase transformations in solids \cite{aes,cagny}, and the stick-slip on rigid substrate \cite{giogio}.
This technique complements the more classical methods used to study the behaviour of physical systems with multiple stable and metastable states \cite{prados1,prados2,tommasi,benic}.
In the context of fracture, the prototypical models here proposed are discrete and based on quasi-one-dimensional lattices composed of breakable and unbreakable bonds. 
While the unbreakable springs serve to distribute the forces in the system, thus describing material elastic energy, the breakable springs are useful in mimicking the fracture propagation.
It is important to note in this context that the role of discreteness in fracture models has already been highlighted in different studies \cite{dis1,dis2,dis3,dis4}. 

In the first  model proposed here ({\it elasto-fragile model}), developed to describe temperature effects in brittle fracture, each breakable spring can be in one of two states, intact or broken, depending on its extension.
Conversely, in the second proposed model ({\it softening-fracture model}), useful to describe temperature effects in brittle-to-ductile transitions, each breakable spring can be in one of three states, intact, softened or broken, depending again from the spring extension. 
In our model, the softened state represents an intermediate configuration where the elastic constant of the spring is smaller than that of the intact spring, but it is still not zero as instead assumed for the broken configuration. 
This intermediate state represents here the counterpart of the material behavior of the cohesive zone introduced in the Dugdale-Barenblatt model, having also the role of introducing an internal length scale.  Both proposed models are approached by calculating the exact partition function, by an approximation obtained for large values of the number $N$ of breakable springs, and, finally by the analysis of the thermodynamic limit. 
This multifaceted treatment allows us to state that both models exhibit a critical behavior with an associated phase transition, whose mechanical implications are thoroughly discussed.
 In our opinion the results are particularly useful for the interpretation of failure processes in micro- and nano-systems, where the effect of temperature is typically studied experimentally and through molecular dynamics simulations \cite{pennings,cai1,weber,cai2,Srolovitz,mao,maeder,saha,zhu,rubber,coarse,dna}.
We want to emphasize that our models, being discrete and addressed to the study of thermal fluctuations, neglect important aspects related to the distribution of elastic fields around the fracture. This is consistent with the fact that they are not developed to replace classical models of continuum mechanics but rather to provide new elements to improve and complement them.

The paper is structured as follows. In Section \ref{simple}, we introduce the first  model for the brittle fracture and we apply the tools of statistical mechanics to eventually obtain exact results. 
Then,  in Section \ref{asybrittle}, we obtain an approximate analytic solution for systems with a large number $N$ of breakable springs, and in Section \ref{thermo}, we study the thermodynamic limit with $N\to\infty$. This allows for a generalization of Griffith's criterion that takes temperature into account by means of a critical behavior.
Concerning the model with the softening mechanism, we introduce its structure and we elaborate its formalism in Section \ref{modelsoft}. Further, we obtain its asymptotic behavior for large values of $N$ in Section \ref{asysoft}, and we study the thermodynamic limit with $N\to\infty$ in Section \ref{thermosoft}. Here we obtain the closed form expression for the brittle-to-ductile transition temperature and describe the corresponding complex critical scenario.

\section{Elasto-fragile model}
\label{simple}

\begin{figure}[t]
    \centering
    \includegraphics[scale=1]{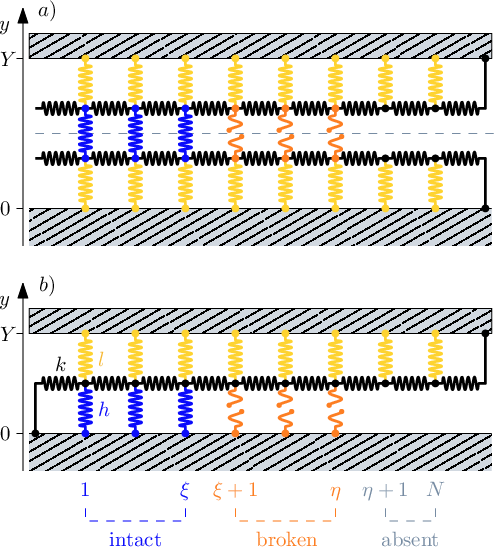}
    \caption{Panel (a): scheme of a crack propagating within an arbitrary crystal lattice. Panel (b):  reduced scheme of the elasto-fragile model, based on symmetry assumption. The central horizontal chain (colored in black) is composed of $N+1$ linear elastic springs of elastic constant $k$. The nodes of this chain are connected to the top layer of the system (at $y=Y$) with $N$ vertical linear elastic springs of elastic constant $l$ (colored in yellow or light gray). Moreover, the first $\eta$ nodes ($i=1,\dots,\eta$) are also linked to the bottom layer (at $y=0$) through $\eta$ vertical breakable springs of elastic constant $h$ (intact in blue or dark gray; broken in orange or intermediate gray). We underline that the first node ($i=0$) and the last one ($i=N+1$) are anchored to the bottom and the top layer, respectively. Hence, the first and the last shear springs fix the direction of the crack propagation from the right to the left of the system.}
    \label{fig_FRA:Fracture_scheme}
\end{figure}

\begin{figure} [t]
    \centering
    \includegraphics[scale=1]{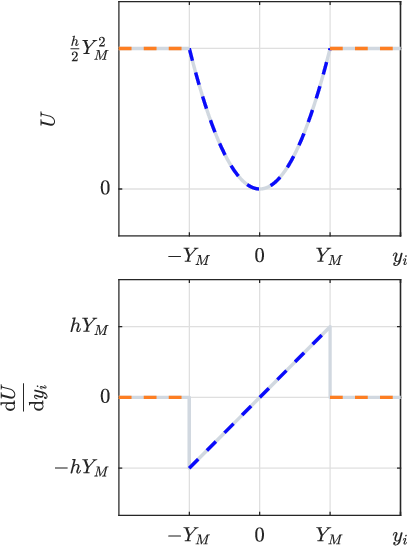}
    \caption{Potential energy of a single breakable spring with elastic constant $h$ (top panel) and corresponding force (bottom panel). The quantity $Y_M$ is the elongation after which the spring breaks, resulting in a force equal to zero.  }
    \label{fig:FRA:Breakable_energy}
\end{figure}

We introduce here a discrete model that helps us to better understand the effect of thermal fluctuations on the brittle fracture processes in solid materials.
As shown in Fig.\ref{fig_FRA:Fracture_scheme}, the model consists in a network of springs with different elastic constants arranged in a quasi-one-dimensional lattice, aimed at reproducing a mode I fracture geometry \cite{giordaRCC}.
Based on symmetry assumptions, we reduce the scheme in Fig.\ref{fig_FRA:Fracture_scheme}(a) to the one in Fig.\ref{fig_FRA:Fracture_scheme}(b).
The structure in Fig.\ref{fig_FRA:Fracture_scheme}(b) connects a fixed substrate, at $y=0$, with a rigid top layer that can be placed at different heights $y=Y$ (isometric conditions within the Helmholtz ensemble).
More in detail, see Fig.\ref{fig_FRA:Fracture_scheme}(b), this structure is composed of a series of $N+1$ springs, with elastic constant $k$, linked together to form an horizontal  chain (colored in black).
The left end-side of the chain, at $i=0$, is attached to the bottom fixed substrate at $y=0$ while the other end, located at $i=N+1$, is attached to the top layer at $y=Y$.
The inner points of the chain, identified by $i=1,\dots ,N$, are individually linked to the top layer through $N$ vertical springs with elastic constant $l$ (colored in yellow or light gray), mimicking the elasticity of the upper half plane. Moreover, the first  ones ($i=1,\dots ,\eta$), are linked to the bottom layer through $\eta$ breakable springs (intact in blue or dark gray; broken in orange or intermediate gray).
We assume that it exists an elongation threshold $Y_M$ after which the potential energy for a breakable spring is constant and the resulting elastic force is zero (broken state), see Fig.\ref{fig:FRA:Breakable_energy}.
The behavior of a single breakable spring corresponds to an elastic constant $h$ when its elongation $y_i$ does not exceed the  threshold $Y_M$ (see Fig.\ref{fig:FRA:Breakable_energy}).
When $\lvert y_i\rvert>Y_M$ the potential energy is constant and then the resulting force is zero (see Fig.\ref{fig:FRA:Breakable_energy}).

We remark that we adopted here simple piecewise linear constitutive equations for the springs of the system. We wish to mention that several important results have been obtained for the crack propagation in nonlinear materials described by power-law stress-strain behavior \cite{nonlin1,nonlin2,nonlin3}. However, from one side it is difficult to combine the thermal analysis with nonlinear materials, and from the other side the nonlinear phenomena are rather limited in nanoscopic systems \cite{cai2,weber}. It is also worth mentioning that due to the discrete quasi-one-dimensional structure  of our model, it is not possible to find here the results concerning the stress singularities at the crack tip and the calculation of the corresponding stress intensity factor (this is true for both the elasto-fragile model and the softening-fracture model) \cite{giordaPRB,giordaRCC}.

We analyze the fracture behavior of the proposed model in the framework of equilibrium statistical mechanics, introducing a temperature $T$ of an embedding thermal bath.  As previously anticipated, we investigate its behavior by adopting isometric conditions corresponding to the Helmholtz ensemble \cite{jpcold,gibbs,weiner}.
We make the assumption that, during the system extension (i.e., increasing $Y$), the system is composed of a segment with $\xi$ intact elements on the left side of the system, and of a segment with $\eta-\xi$ broken elements on the right. As a result, the system evolution is characterized by the propagation of  a single interface between intact and broken springs,  regulated by the assigned traction conditions and by the temperature.  
This hypothesis (known as single domain wall assumption) simplifies the calculations and makes it possible to analytically derive the partition function and, thus, the important macroscopic physical quantities. The same hypothesis is considered in the classical continuum fracture models recalled in the introduction.
The configurations previously described can be summarized by the relation $1\leq\xi\leq\eta\leq N$, where $\xi$ represents the domain wall or interface position.
We remark that the last $N-\eta$ sites of the chain are always considered disconnected from the bottom layer in order to simulate a  possible existing initial  fractured domain. This is coherent with the assumption of an initial crack extension in the classical Griffith criterion, which is the milestone of the linear elastic fracture mechanics \cite{griffith,pugno}. 

Based on previous key premises, the total energy of the system is
\begin{eqnarray}
\nonumber
    \Phi_H(y_1,...,y_N,\xi) &=& \sum_{i=0}^N\frac{k}{2}(y_{i+1}-y_i)^2 + \sum_{i=1}^N\frac{l}{2}(Y-y_i)^2 \\
    \label{eq:fra:hamiltonian}
    &&+ \sum_{i=1}^\xi\frac{h}{2}y_i^2 + \sum_{i=\xi+1}^\eta\frac{h}{2}Y_M^2,
\end{eqnarray}
with the boundary conditions $y_0=0$ and $y_{N+1}=Y$. Here, the variables $y_i$ represent the vertical coordinates of the lattice points while $\xi$ assignes the discrete interface position. 
Observe that the first addend of Eq.\eqref{eq:fra:hamiltonian}, proportional to the elastic constant $k$, is the energetic contribution of the shear unbreakable springs. 
We remark that the use of the shear springs with elastic constant $k$ is an approximation valid under the small deformation assumption (in our models the central nodes can only move vertically).
The second addend, proportional to the elastic constant $l$, is the contribution of the vertical  unbreakable springs that connect the upper layer to the inner lattice points.
Finally, the energetic contribution proportional to the elastic constant $h$, regarding the breakable springs, is split into two addends: the first one corresponds to the unbroken springs (going from $i=1$ to $i=\xi$) and the second one corresponds to the broken ones (from $i=\xi+1$ to $i=\eta$).
Again, the assumption of linear elastic springs, as typical in linear elastic fracture mechanics, allow for a proper description of fracture effects under the hypothesis of small strains. Such a simplification is crucial for the following analytical treatment.
The energy contribution associated with broken bonds corresponds to the surface energy of the two exposed sides of the fracture, originally introduced by Griffith in the overall energy balance and eventually yielding his classical stability criterion \cite{griffith}.

The energy in Eq.\eqref{eq:fra:hamiltonian} can be rewritten as
\begin{equation}
    \begin{aligned}
            \Phi_H =& \frac{k}{2}\left[\sum_{i=1}^{N}\left(2+\frac{l}{k}\right)y_i^2 + \sum_{i=1}^{\xi}\frac{h}{k}y_i^2 - \sum_{i=1}^{N-1}y_{i+1}y_i - \sum_{i=1}^{N-1}y_iy_{i+1}\right]\\
            +&kY\left[-\sum_{i=1}^N\frac{l}{k}y_i-y_N\right] + \frac{1}{2}kY^2 + \frac{1}{2}lNY^2 + \frac{1}{2}hY_M^2(\eta-\xi).
    \end{aligned}
    \label{enenene}
\end{equation}
We can introduce the following $N$-component vectors
\begin{align}
\label{vet1}
    \vec{y}&=    \left( y_1 , y_2 , \dots , y_N\right) ,
   \\
   \label{vet2}
    \vec{v}&=   \left(   \beta , \dots , \beta , 1+\beta\right) ,
    \end{align}
and the tridiagonal matrix
\begin{equation}
    \label{eq:fra:tridiagonal_matrix}
    \mathcal{A}=\left[
    \begin{array}{ccccc}
            a_1    & -1     & 0      & \dotsm & 0      \\
        -1     & \ddots & \ddots & \ddots & \vdots \\
        0      & \ddots &  \ddots      &   \ddots     & 0      \\
        \vdots & \ddots &     \ddots   &    \ddots    & -1     \\
        0      & \dotsm & 0      & -1     & a_N
    \end{array} \right],
\end{equation}
where the diagonal components $a_i$ are defined as follows
\begin{equation}
    a_i =
    \begin{cases}
        2+\alpha,&\text{ if }\,1\leq i\leq \xi,\\
        2+\beta,&\text{ if }\,\,\xi+1\leq i\leq N,
    \end{cases}
    \label{aaaiii}
\end{equation}
with
\begin{equation}
            \alpha = \frac{l+h}{k},\,\,\,\,\,\,
        \beta = \frac{l}{k},
    \end{equation}
measuring extension versus shear springs stiffness of the lattice.
Adopting the matrix $\mathcal{A}$ and the vectors $\vec{y}$ and $\vec{v}$, we can write Eq.(\ref{enenene}) as
\begin{equation}
    \Phi_H = \frac{1}{2}k\vec{y}\cdot\mathcal{A}\vec{y} - kY\vec{v}\cdot\vec{y} + \frac{1}{2}kY^2 + \frac{1}{2}lNY^2 + \frac{1}{2}hY_M^2(\eta-\xi).
\end{equation}
This new compact expression is more suitable to evaluate the partition function of the system, defined by
\begin{equation}
    Z_H(Y)=\sum_{\xi=0}^\eta\int_{\mathbb{R}^N} e^{-\frac{\Phi_H}{K_BT}}\mathrm{d} \vec{y}.
\end{equation}
Here, we have integrated the continuous coordinates $y_i$ and we have summed over the discrete or spin variable $\xi$ identifying the interface position.
We have
\begin{equation}
    \label{eq:fra:part_fun}
    Z_H(Y)=\sum_{\xi=0}^\eta\exp{\left[ -\frac{kY^2}{2K_BT}-\frac{lNY^2}{2K_BT}-\frac{hY_M^2(\eta-\xi)}{2K_BT}\right] }\mathcal{I}_\xi,
\end{equation}
where  
\begin{equation}
    \mathcal{I}_\xi = \int_{\mathbb{R}^N}\exp{\left( -\frac{k}{2K_BT}\vec{y}\cdot\mathcal{A}\vec{y}+\frac{kY}{K_BT}\vec{v}\cdot\vec{y}\right) }\mathrm{d} \vec{y}.
\end{equation}
By using the classical Gaussian integral,
\begin{equation}
    \label{eq:fra:Gauss_matrix}
    \int_{\mathbb{R}^N}e^{-\vec{y}\cdot\mathcal{M}\vec{y}}e^{\vec{w}\cdot\vec{y}}\mathrm{d} \vec{y}=\sqrt{\frac{\pi^N}{\det \mathcal{M}}}e^{\frac{1}{4}\vec{w}\cdot\mathcal{M}^{-1}\vec{w}},
\end{equation}
which is valid for a positive definite symmetric matrix $\mathcal{M}$ (as
can be shown for the tridiagonal matrix $\mathcal{A}$) and for any vector $\vec{w}$, we get
\begin{equation}
    \mathcal{I}_\xi =\sqrt{\frac{(2\pi K_BT)^N}{k^N\det \mathcal{A}}}\exp{\left( \frac{kY^2}{2K_BT}\vec{v}\cdot\mathcal{A}^{-1}\vec{v}\right) }.
\end{equation}
Thus, the partition function can be written as
\begin{equation}
    \begin{aligned}
        Z_H(Y)=&\sum_{\xi=0}^\eta\exp{\left[ -\frac{kY^2}{2K_BT}-\frac{lNY^2}{2K_BT}-\frac{hY_M^2(\eta-\xi)}{2K_BT}\right] }\\
        &\times\sqrt{\frac{(2\pi K_BT)^N}{k^N\det\mathcal{A}}}\exp{\left( \frac{kY^2}{2K_BT}\vec{v}\cdot\mathcal{A}^{-1}\vec{v}\right) }.
    \end{aligned}
    \label{ZZZ}
\end{equation}
Since $\mathcal{A}$ depends on  $\xi$, see Eqs.(\ref{eq:fra:tridiagonal_matrix}) and (\ref{aaaiii}), both $\mathcal{A}^{-1}$ and  $\det\mathcal{A}$ depend on  $\xi$ in the sum of Eq.(\ref{ZZZ}).
In Appendix \ref{appendixa}, we discuss an efficient method to determine $\mathcal{A}^{-1}$ and  $\det\mathcal{A}$ for a tridiagonal matrix. This method will be used to obtain asymptotic expressions, useful to study the system behavior for large values of $N$ (and for the thermodynamic limit). 
By introducing the quantity
\begin{equation}
\label{eq:fra:q}
    q=1+\beta N-\vec{v}\cdot\mathcal{A}^{-1}\vec{v},
\end{equation}
we can write the partition function in the  form
\begin{equation}
\label{eq:fra:HpartfunQ}
    Z_H(Y)=\sum_{\xi=0}^\eta\sqrt{\frac{(2\pi K_BT)^N}{k^N\det \mathcal{A}}}\exp{\left[ \frac{-hY_M^2(\eta-\xi)-kY^2q}{2K_BT}\right] }.
\end{equation}

We  can now  evaluate the expected values of macroscopic quantities.
For example, expectation value of the force applied to the system is \cite{jpcold,gibbs,weiner}
\begin{equation}
\label{deri}
    \langle f \rangle = -K_BT\frac{1}{Z_H}\frac{\partial Z_H}{\partial Y},
\end{equation}
resulting in
\begin{align}
    \label{eq:fra:f-e_Helm2}
    \langle f \rangle=&\frac{\sum_{\xi=0}^\eta(\det\mathcal{A})^{-\frac{1}{2}}\exp{\left[ \frac{-hY_M^2(\eta-\xi)-kY^2q}{2K_BT}\right]}q }{\sum_{\xi=0}^\eta(\det\mathcal{A})^{-\frac{1}{2}}\exp{\left[ \frac{-hY_M^2(\eta-\xi)-kY^2q}{2K_BT}\right]} } k\, Y .
    \end{align}
    This expression gives a physical interpretation to the expectation value $\langle q \rangle$ of the variable $q$. Since $\langle f \rangle=k\langle q \rangle Y$, the quantity  $k \langle q \rangle$ represents the effective stiffness $\langle f \rangle/Y$ of the system.
Using a similar analysis, it is possible to obtain the average value of the number of unbroken bonds, which is given by
 \begin{align}
    \label{eq:fra:xi_Helm2}
    \langle \xi \rangle=&\frac{\sum_{\xi=0}^\eta(\det\mathcal{A})^{-\frac{1}{2}} \exp{\left[ \frac{-hY_M^2(\eta-\xi)-kY^2q}{2K_BT}\right]}\xi }{\sum_{\xi=0}^\eta(\det\mathcal{A})^{-\frac{1}{2}}\exp{\left[ \frac{-hY_M^2(\eta-\xi)-kY^2q}{2K_BT}\right] }}.
\end{align}

\begin{figure}
    \centering
    \includegraphics[scale=1]{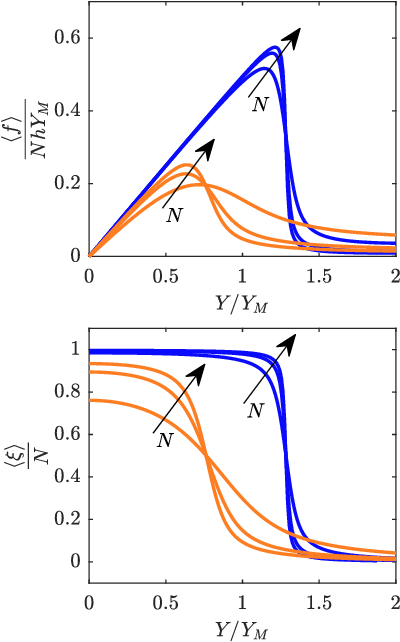}
    \caption{Behavior of the brittle-fracture model with a variable number of units ($N=\{50, 125, 200\}$ as indicated by arrows) and a variable thermal to elastic energy ratio $K_BT/(hY_M^2)=0.5$ (blue or dark gray), and $K_BT/(hY_M^2)=2$ (orange or light gray). The dimensionless quantities $\langle f\rangle/(NhY_M)$ (top panel) and $\langle \xi\rangle/N$ (bottom panel) are represented versus the dimensionless parameter $Y/Y_M$, where, in both cases, $l/h=k/h=1$. We also fixed $\eta=N$, which means there are no missing or broken elements in the initial configuration.}
    \label{fig:FRA:exact_fracture}
\end{figure}

In Fig.\ref{fig:FRA:exact_fracture}, we show the main effects on the fracture behavior of both temperature $T$ and discreteness parameter $N$.
The main effect that can be observed is related to temperature, which is able to shift the value of the extension corresponding to the fracture of the system. 
In particular, we may observe that, as typically experimentally observed, the higher the temperature, the lower the force and the extension required to induce fracture. The model then  predicts a thermally activated fracture phenomenon. 
 As described in the following, in the thermodynamic limit (i.e., for $N\to\infty$), this behavior can be theoretically interpreted as a phase transition. 
We can observe since now (see Fig.\ref{fig:FRA:exact_fracture}) that, as the discreteness parameter $N$ increases, the force-displacements curves become sharper, increasing the brittleness of the system. In summary, the model exhibits a temperature dependent brittle behavior, which can be thoroughly described by analytic expressions for large values of $N$, obtained in the following Section.

\section{Asymptotic behavior of the elasto-fragile model}
\label{asybrittle}
\begin{figure}
    \centering
    \includegraphics[scale=1]{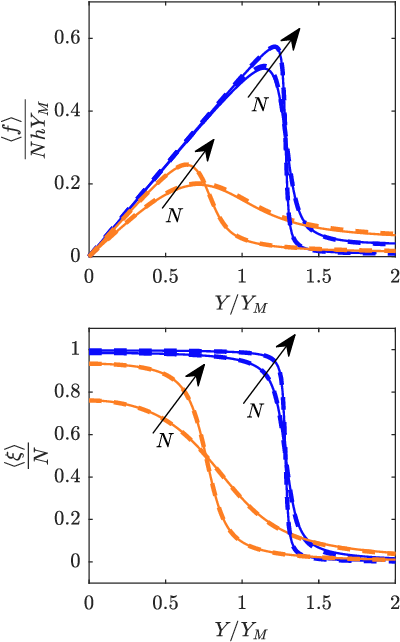}
    \caption{Comparison between the approximated quantities given by Eqs.(\ref{fapp}) and (\ref{xiapp}) (dashed lines) and the corresponding exact results in Eqs.(\ref{eq:fra:f-e_Helm2}) and (\ref{eq:fra:xi_Helm2}) (continuous lines) for $\langle f\rangle/(NhY_M)$ (top panel) and $\langle \xi\rangle/N$ (bottom panel) versus the dimensionless extension $Y/Y_M$. The thermal to elastic energy ratio $K_BT/(hY_M^2)$ is set to $0.5$ (blue or dark gray curves) and $2$ (orange or light gray curves). The total number of units is set to $N=50$ and $N=200$ (see arrows). The dimensionless quantities are set to $l/h=1$ and $k/h=1$. We also considered $\eta=N$.}
    \label{fig:FRA:largeN_fracture}
\end{figure}

Eqs.(\ref{eq:fra:f-e_Helm2}) and (\ref{eq:fra:xi_Helm2})  determine the expected value of $\langle f\rangle$, the force applied to the system, and the average value $\langle \xi\rangle$ of the number of intact bonds  as functions of both the assigned displacement $Y$ and of the temperature $T$.
Here, to give a clearer physical interpretation of such results, we obtain analytical approximated relations, effective in the case of large values of $N$. 
In particular, for large values of $N$, we have  (see Appendix \ref{appendixb})
\begin{equation}
    q \sim \frac{lh}{l+h}\frac{\xi}{k} + \epsilon,
    \label{qqtext}
\end{equation}
where
\begin{equation}
    \epsilon = \frac{\sqrt{\beta^2+4\beta}-\beta}{2}>0
    \label{eptext}
\end{equation}
and
\begin{equation}
    \det\mathcal{A} \sim \tau_\alpha^\xi\tau_\beta^{N-\xi},
    \label{dettext}
\end{equation}
with 
\begin{equation}
        \tau_s = \frac{2+s+\sqrt{s^2+4s}}{2}, \hspace{0.5 cm} s=\alpha, \beta.\\
\label{tautau}
\end{equation}
 By using Eq.(\ref{eq:fra:f-e_Helm2}), we obtain the following asymptotic expression of the force-extension relation
\begin{equation}
    \langle f \rangle\sim \frac{\sum_{\xi=0}^\eta\exp{\left( -\frac{\xi}{2}\ln\delta+\frac{hY_M^2\xi}{2K_BT}-\frac{Y^2}{2K_BT}\frac{lh}{l+h}\xi\right) }q}{\sum_{\xi=0}^\eta\exp{\left( -\frac{\xi}{2}\ln\delta+\frac{hY_M^2\xi}{2K_BT}-\frac{Y^2}{2K_BT}\frac{lh}{l+h}\xi\right) }}k\,Y,
\end{equation}
where we introduced $\delta = \tau_\alpha/\tau_\beta$. 
After defining
\begin{equation}
    z =\exp\left( -\frac{1}{2}\ln\delta+\frac{hY_M^2}{2K_BT}-\frac{Y^2}{2K_BT}\frac{lh}{l+h}\right) ,
    \label{zzzz}
\end{equation}
and adopting similar considerations for the quantity $\langle \xi\rangle$, we may easily deduce 
\begin{align}
\label{a111}
    \langle f \rangle \sim &\left ( k \epsilon+ \frac{lh}{l+h}\frac{\sum_{\xi=0}^\eta \xi z^\xi}{\sum_{\xi=0}^\eta z^\xi}\right ) \, Y ,\\
    \label{a222}
    \langle \xi \rangle \sim & \frac{\sum_{\xi=0}^\eta \xi z^\xi}{\sum_{\xi=0}^\eta z^\xi},
\end{align}
for large values of $N$.
Combining Eqs.(\ref{a111}) and (\ref{a222}), we find
\begin{align}
\label{fapp}
    \langle f \rangle \sim \left ( k\epsilon + \frac{lh}{l+h}\langle \xi \rangle\right ) \, Y .
\end{align}
So, if we calculate the expectation value of intact elements $\langle \xi \rangle$, we also deduce the force-extension relation $\langle f \rangle$-$Y$.
To do this in explicit form, we have to evaluate the sums that appear in the expression of $\langle \xi \rangle$, i.e.
\begin{align}
    \label{eq:fra:sums_1}
    \sum_{\xi=0}^\eta \xi z^\xi &= \frac{z\left[1-(\eta+1)z^\eta+\eta z^{\eta+1}\right]}{(1-z)^2},\\
    \label{eq:fra:sums_2}
    \sum_{\xi=0}^\eta z^\xi     &= \frac{1-z^{\eta+1}}{1-z},
\end{align}
where we adopted the variable $z$ defined in Eq.(\ref{zzzz}). To conclude, we write
\begin{align}
        \label{xiapp}
    \langle \xi \rangle \sim \frac{1-(\eta+1)z^\eta+\eta z^{\eta+1}}{1-z^{\eta+1}}\frac{z}{1-z},
\end{align}

These results represent the approximated expressions for the response of the system under isometric condition for large values of $N$.
As shown in Fig.\ref{fig:FRA:largeN_fracture} (for $\eta=N$), we can observe that the approximations given in Eqs.(\ref{fapp}) and (\ref{xiapp}) (dashed lines in the figure) are in perfect agreement with the exact results (continuous lines), previously obtained in Eqs.(\ref{eq:fra:f-e_Helm2}) and (\ref{eq:fra:xi_Helm2}). 
We remark that the agreement is very good for different temperatures and for different (large) values of $N$. These approximated results are particularly useful to study the thermodynamic limit or, equivalently, to study the limiting case for  $N\to\infty$, as discussed below.

\section{Thermodynamic limit of the elasto-fragile model}
\label{thermo}

We perform now the limit for $N\to\infty$. Since previous asymptotic results in Eqs.(\ref{fapp}) and (\ref{xiapp}) depend on powers of $z$, we study the inequality $z>1$. To verify this condition we need to set the argument of the exponential in Eq.(\ref{zzzz}) larger than zero, obtaining the following condition on $Y$
\begin{equation}
\label{absy}
   \lvert Y \rvert \leq \sqrt{\frac{l+h}{lh}\left[hY_M^2-K_BT\ln \delta\right]}\triangleq Y_s,
\end{equation}
where we introduced the critical extension $Y_s$, the physical interpretation of which will be given below. 
We also define a critical temperature $T_c$ for the system through the condition $hY_M^2-K_BT_c\ln \delta = 0$ (see Eq.(\ref{absy})) that, once solved for $T_c$, gives
\begin{equation}
    \label{eq:fra:Tc}
    T_c = \frac{hY_M^2}{\displaystyle{K_B\ln\left(\frac{2+\frac{l+h}{k}+\sqrt{\left(\frac{l+h}{k}\right)^2+4\frac{l+h}{k}}}{2+\frac{l}{k}+\sqrt{\left(\frac{l}{k}\right)^2+4\frac{l}{k}}}\right)}},
\end{equation}
where we used $\delta = \tau_\alpha/\tau_\beta$.
This is a specific value of the temperature, depending on the main material parameters of the system, which corresponds to a phase transition, as discussed below.
For the moment, we can write
\begin{equation}
    z>1 \iff \lvert Y \rvert \leq\sqrt{\frac{l+h}{l}Y_M^2\left(1-\frac{T}{T_c}\right)}\triangleq Y_s
\end{equation}
so that $Y_s=0$ when $T=T_c$.

\begin{figure}[t]
    \centering
    \includegraphics[scale=1]{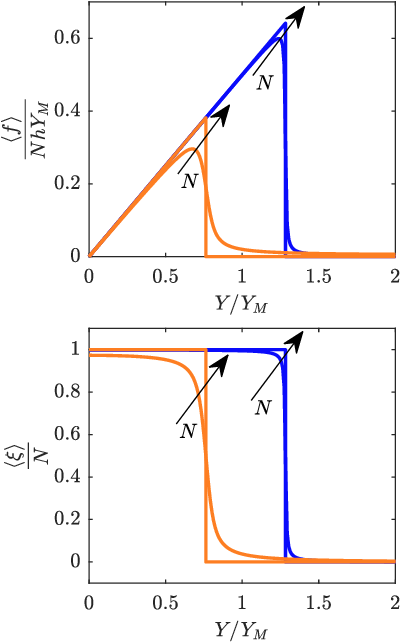}
    \caption{Comparison between the thermodynamic limit obtained for $N\to\infty$ and the approximations obtained with large values of $N$ (see arrows). We plotted $\langle f\rangle/(NhY_M)$ (top panel) and $\langle \xi\rangle/N$ (bottom panel) versus the dimensionless extension $Y/Y_M$. The thermal to elastic energy ratio is set to $K_BT/(hY_M^2)=0.5$ (in blue or dark gray) and to $K_BT/(hY_M^2)=2$ (in orange or light gray) while the total number of units for the large $N$ approximation is set to $N=500$. The dimensionless quantities are set to $l/h=1$ and $k/h=1$. Moreover, we adopted $\eta=N$ corresponding to $\phi=0$.}
    \label{fig:FRA:infN_fracture}
\end{figure}

We first analyze the limit for $N\to\infty$ of the average fraction of intact elements
\begin{equation}
    \begin{aligned}
        \frac{\langle \xi \rangle}{N} =&\frac{1-(\eta+1)z^\eta+\eta z^{\eta+1}}{1-z^{\eta+1}}\frac{z}{1-z}\frac{1}{N}\\
    =&\frac{1-z^{\eta+1}+(\eta+1)(z^{\eta+1}-z^\eta)}{1-z^{\eta+1}}\frac{z}{1-z}\frac{1}{N}\\
    =&\left(\frac{1}{N}+\frac{\eta+1}{N}\frac{1-z}{z-z^{-\eta}}\right)\frac{z}{1-z}.
    \end{aligned}
\end{equation}
Now, we consider $\eta=N-M$, where $M=\phi N$ is the number of initially absent breakable springs  (initial fractured domain). Here, $\phi$ is the percentage of initially absent breakable springs over the total number $N$ of elements.
The fraction of intact elements is determined as follows
\begin{equation}
    \lim_{N\to\infty}\frac{\langle \xi \rangle}{N}=
    \begin{cases}
    1-\phi&\quad\text{if }z>1 \mbox { or } \lvert Y \rvert < Y_s,\\
    0&\quad\text{if }z<1 \mbox { or } \lvert Y \rvert > Y_s.
    \end{cases}
\end{equation}
This means that all elements are broken simultaneously (brittle fracture) when $Y=Y_s$.

\begin{figure}[t]
    \centering
    \includegraphics[scale=1]{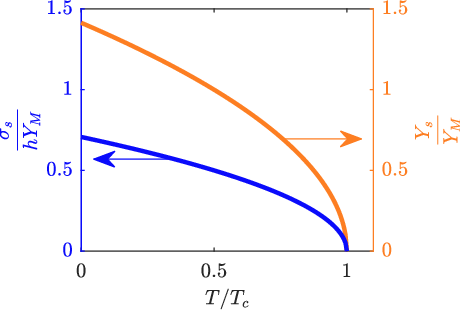}
    \caption{ Fracture extension $Y_s/Y_M$ (in orange or light gray) and fracture strength $\sigma_s/(hY_M)$ (in blue or dark gray) versus the reduced temperature $T/T_c$. All quantities are written in dimensionless form. We can observe that both quantities present a critical behavior corresponding to a phase transition for $T=T_c$. We adopted $l/h=1$, $k/h=1$, and $\phi=0$.}
    \label{fig:fra:InfN_Critic}
\end{figure}

For the stress $\langle f \rangle/N$ (density of force over the number of bonds), we get
\begin{equation}
    \lim_{N\to\infty}\frac{\langle f \rangle}{N}=
    \begin{cases}
    \frac{lh}{l+h}(1-\phi) Y&\quad\text{if }z>1 \mbox { or } \lvert Y \rvert < Y_s,\\
    0&\quad\text{if }z<1 \mbox { or } \lvert Y \rvert > Y_s.
    \end{cases}
\end{equation}
Thus (see Fig.\ref{fig:FRA:infN_fracture}), after an initial linear behavior, the stress collapses to zero at the  extension threshold $Y_s$.
Together with the  extension threshold $Y_s$, we can therefore introduce the  stress threshold $\sigma_s$ as follows
\begin{align}
    Y_s =& Y_M\sqrt{\frac{l+h}{l}\left(1-\frac{T}{T_c}\right)},\\
    \label{eq:fra:yieldStrength}
    \sigma_s =& hY_M(1-\phi)\sqrt{\frac{l}{l+h}\left(1-\frac{T}{T_c}\right)}.
\end{align}
In other words, $\sigma_s$ is the value of ${\langle f \rangle}/{N}$ in correspondence of $Y=Y_s$. We can say that $Y_s$ is the fracture extension while $\sigma_s$ is the fracture strength inducing the breaking process.

The behavior of the system is shown in Fig.\ref{fig:FRA:infN_fracture}, where the dimensionless quantities  $\langle f\rangle/(NhY_M)$ and $\langle \xi\rangle/N$ are represented versus the dimensionless  extension $Y/Y_M$. 
We compared here the response for a large value of $N$ and the thermodynamic limit. In this limit, the breaking is fully brittle with temperature dependent fracture extension and stress. 
Thus, as typical in collective phenomena of complex systems, although each breakable spring has a temperature-independent breakage threshold, the overall system exhibits a temperature-dependent fracture point due to the interactions between the springs and the thermal bath. 
In particular, the system undergoes a phase transition for $T=T_c$, with both fracture extension and stress decreasing to zero at $T=T_c$ when the system breaks without any external mechanical actions. 
This is described in Fig.\ref{fig:fra:InfN_Critic}, where we plot the dimensionless fracture extension $Y_s/Y_M$  and the dimensionless fracture strength $\sigma_s/(hY_M)$ versus the temperature ratio $T/T_c$. This critical behavior corresponds to a classical second order phase transition.
We also remark that Eq.\eqref{eq:fra:yieldStrength} represents an extension of the Griffith criterion of the linear elastic fracture mechanics, accounting for the additional effects of temperature.
From Fig.\ref{fig:FRA:Breakable_energy}, we see that the energy necessary to break an element is given by $hY_M^2/2$
and then the Griffith surface energy density  $\gamma_s$ is proportional to $hY_M^2/2$. Equivalently, $Y_M$ is proportional to $\sqrt{\gamma_s}$ and therefore it is easily seen that the fracture strength given in Eq.(\ref{eq:fra:yieldStrength}) is proportional to $\sqrt{\gamma_s}$, exactly as in Griffith's criterion \cite{griffith}.
Moreover, at constant temperature it is well seen that the breaking strength decreases if $\phi$ increases, which is exactly what the Griffith's criterion states \cite{griffith}. 
This means that, if the initial system is degraded, a smaller force is required to continue its mechanical degradation.
Of course, our version is quantitatively different from the original one because of the simplified geometry we used.
In particular, we do not consider the exact elastic energy distributed over the deformed continuum due to fracture.
Our model, however, introduces thermal effects in brittle fracture and, in particular, shows the phase transition at the critical temperature $T_c$ given in Eq.\eqref{eq:fra:Tc}.

\begin{figure}
    \centering
    \includegraphics[scale=1]{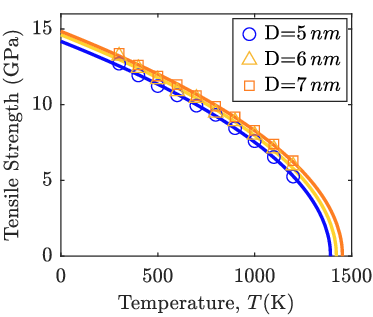}
    \caption{Tensile strength of [110]-oriented silicon nanowires as a function of temperature for wires with different diameters: comparison between molecular dynamics simulations results \cite{cai2}  (symbols) and our theory given by Eq.(\ref{eq:fra:yieldStrength}) (continuous lines). The parameters used are reported in the main text.}
    \label{brittle-fit}
\end{figure}

To show the effectiveness of the obtained results, we analyze the temperature dependent fracture behavior of [110]-oriented silicon nanowires \cite{cai2}.
In Fig.\ref{brittle-fit}, we compare the theoretical fracture force given by  Eq.(\ref{eq:fra:yieldStrength}) with the molecular dynamics results discussed in Ref.\cite{cai2}.
We observe that the theory well predicts the brittle fracture behavior of the nanowires both temperature- and diameter-wise (see Fig.4(b) of Ref.\cite{cai2}). 
In the figure the theoretical force in Eq.(\ref{eq:fra:yieldStrength}) has been divided by the area $\mathsf{S}$ pertaining to each breakable spring in order to obtain the stress $\sigma=\left\langle f\right\rangle /(N\mathsf{S})=\sigma_s/\mathsf{S}$. 
The strain has been determined as $\varepsilon=Y/\ell$, where $\ell$ is the characteristic lengthscale induced by the crystal structure. For all curves we adopted the parameters $Y_M = 1.78\times 10^{-11}$m, $\mathsf{S}= 2.27\times 10^{-21}$m$^2$, $\ell=1.82\times 10^{-10}$m, $k   = 88.4$N/m, and $K_B = 1.38\times 10^{-23}$J/K.
Moreover, for the blue curve (or dark gray, $D = 5$nm) we used $l   = 9.07$N/m and $h   = 2.00$N/m; for the yellow curve (or light gray, $D = 6$nm) we used $l   = 9.54$N/m and $h   = 2.05$N/m; for the orange curve (or intermediate gray, $D = 7$nm) we used $l   = 9.94$N/m and $h = 2.08$N/m.
While most of geometrical parameters were available in the original paper dealing with molecular dynamics simulations, the other physical parameters (in particular the elastic constants), were fitted to correctly reproduce the results. The elastic constants take effective values pertinent to the springs of our lattices and therefore cannot be directly obtained from the data available in the above papers. Interestingly, all the obtained (fitted) values are reasonable and consistent with the underlying physics of the system.
In particular, the fact that $h$ and $l$ increase with the  diameter is consistent with the results of Ref.\cite{cai2}, providing evidence that the nanowires Young modulus $E$ increases with diameter (scale effect). 
This coherence is also quantitative since in our case we have $E=\ell/[(1/l+1/h)\mathsf{S}]$, which assumes the values 130GPa, 135GPa and 138GPa, for the three diameters 5nm, 6nm and 7nm, in agreement with Fig.4(a) of Ref.\cite{cai2}.        
The good agreement between theory and simulations makes us confident on the applicability of our theory to micro- and nanoscopic systems.

\section{Softening-Fracture model}
\label{modelsoft}
\begin{figure}
    \centering
    \includegraphics[scale=1]{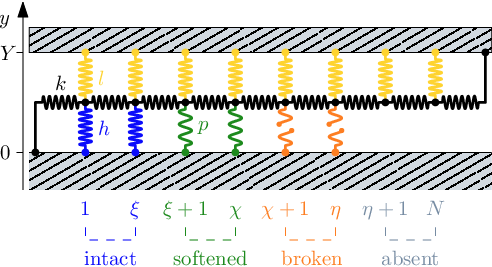}
    \caption{ Scheme of the fracture model with the softening mechanism. The central horizontal chain (colored in black) is composed by  $N+1$ linear springs with elastic constant $k$. The nodes of this chain are connected to the top layer (at $y=Y$) by $N$ vertical linear springs with elastic constant $l$ (colored in yellow or light gray). The first $\eta$  nodes ($i=1,\dots,\eta$) are also linked to the bottom layer (at $y=0$) by $\eta$ vertical softenable and breakable springs with elastic constant $h$ when intact (colored in blue or dark gray), or $p$ when softened (colored in green, springs with less coils). The broken elements are represented in orange and identified by a rupture in the springs. We remark that the first node ($i=0$) and the last one ($i=N+1$) are anchored to the bottom and the top layers, respectively.}
    \label{fig_FRA:Fracture_Scheme_Softened}
\end{figure}

\begin{figure}[t!]
    \centering
    \includegraphics[scale=1]{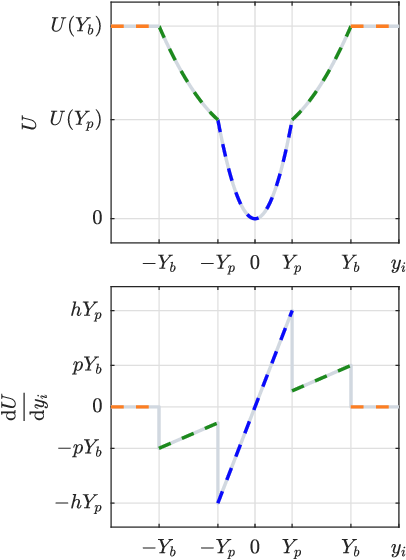}
    \caption{Potential energy of a single softenable and breakable spring of elastic constants $h$ and $p$ (top panel) and corresponding force (bottom panel). We see that $Y_p$ is the elongation after which the spring is weakened or softened, and $Y_b$ is the elongation after which the spring is broken.}
    \label{fig:FRA:Breakable_energy_Soft}
\end{figure}

The previous model, useful to describe brittle fracture, is further generalized here to introduce a material ductile behavior of the elements possibly resulting in a brittle-to-ductile transition.
Specifically, in the same spirit of the Dugdale-Barenblatt model \cite{dugdale,barenblatt}, we introduce a cohesive zone between the elastic and fractured domains of the breakable springs, characterized by two different states before the broken configuration, depending on their extension $y_i$ (see  Fig.\ref{fig_FRA:Fracture_Scheme_Softened}).
More precisely, each breakable spring presents an elastic constant $h$ when its extension is less than the softening point $Y_p$ and a lower elastic constant $p<h$ for larger extensions, until the breaking point corresponding to the extension $Y_b$ is attained and the link is broken (see Fig.\ref{fig:FRA:Breakable_energy_Soft}).
As we can see, each breakable element behaves as a spring of elastic constant $h$ when  $-Y_p\leq y_i\leq Y_p$.
Then, the spring is softened with an elastic constant $p<h$ when $Y_p \leq \lvert y_i \rvert  \leq Y_b$.
After the breaking point $Y_b$, the potential energy is constant and therefore the resulting force is zero.
Thus, the potential energy of a breakable spring is
\begin{equation}
    U(y_i) =
    \begin{cases}
           \frac{1}{2}hy_i^2&\quad\text{if }\lvert y_i \rvert \leq Y_p,\\
           \frac{1}{2}py_i^2+\Delta E&\quad\text{if }Y_p\leq\lvert y_i \rvert \leq Y_b,\\
           \frac{1}{2}pY_b^2+\Delta E&\quad\text{if }\lvert y_i \rvert \geq Y_b.\\
    \end{cases}
\end{equation}
Given the two elastic moduli $h$ and $p$ (with $p<h$) and the energy gap $\Delta E>0$ , we obtain the softening point
\begin{equation}
    Y_p = \sqrt{\frac{2\Delta E}{h-p}},
\end{equation}
which must always satisfy the condition $Y_p<Y_b$. 
Thus $\Delta E+pY_p^2/2=hY_p^2/2$ is the energy necessary to weaken one breakable element of the system, and $\Delta E+pY_b^2/2=hY_p^2/2+p(Y_b^2-Y_p^2)/2$ is the energy necessary to break the element.  This reproduces in the discrete context considered here the Irwin generalization of the Griffith's criterion \cite{Irwin1,Irwin2}.

 The total energy $\Phi_H(y_1,...,y_N,\xi,\chi)$ of the system is
\begin{equation}
    \label{eq:fra:soft_hamiltonian}
    \begin{aligned}
        \Phi_H =& \sum_{i=0}^N\frac{k}{2}(y_{i+1}-y_i)^2 + \sum_{i=1}^N\frac{l}{2}(Y-y_i)^2 + \sum_{i=1}^\xi\frac{h}{2}y_i^2\\
        &+ \sum_{i=\xi+1}^\chi\left(\frac{p}{2}y_i^2+\Delta E\right)+\sum_{i=\chi+1}^\eta\left(\frac{p}{2}Y_b^2+\Delta E\right).
    \end{aligned}
\end{equation}
Here we introduced the position $\xi$ of the interface between intact and softened elements, the position $\chi$ of the interface between softened and fully broken elements and, finally, the position $\eta$ of the interface between fully broken and initially absent elements. 
The value of $\eta$ corresponds to the initial state of the system and is therefore fixed. 
The two interfaces at $\xi$ and $\chi$ can move as a function of temperature and mechanical actions on the system. 
The aim of this section is to study the  (quasi-static) evolution of these interfaces  determining the fracture propagation phenomenon. 
The region  between $\xi$ and $\chi$, characterized by softened elements, identifies the cohesive zone of the rupture phenomenon. When the cohesive zone is absent or negligible, the fracture is brittle; on the other hand, when the cohesive zone is not negligible, the fracture becomes ductile. 
Therefore, as we show in the following, this model  allows to describe the brittle-to-ductile transitions.
As in Section \ref{simple}, we remark that the use of the shear springs with elastic constant $k$ is an approximation valid under the small deformation assumption (in our models the central nodes can only move vertically).

We can rewrite Eq.(\ref{eq:fra:soft_hamiltonian}) as
\begin{equation}
    \begin{aligned}
            \Phi_H =& \frac{k}{2}\left[\sum_{i=1}^{N}\left(2+\frac{l}{k}\right)y_i^2 + \sum_{i=1}^{\xi}\frac{h}{k}y_i^2 + \sum_{i=\xi+1}^{\chi}\frac{p}{k}y_i^2 - 2\sum_{i=1}^{N-1}y_{i+1}y_i\right]\\
            &+kY\left[-\sum_{i=1}^N\frac{l}{k}y_i-y_N\right] + \frac{1}{2}kY^2 + \frac{1}{2}lNY^2\\
            &+\left(\frac{p}{2}Y_b^2+\Delta E\right)(\eta-\chi)+(\chi-\xi)\Delta E.
    \end{aligned}
\end{equation}
As before, to simplify the mathematical structure of the energy function, we introduce the vectors in Eqs.(\ref{vet1}) and (\ref{vet2}), and the tridigonal matrix in Eq.(\ref{eq:fra:tridiagonal_matrix}), where the diagonal elements $a_i$ are now defined as follows
\begin{equation}
    a_i =
    \begin{cases}
        2+\alpha&\text{ if }\,1\leq i\leq \xi,\\
        2+\gamma&\text{ if }\,\,\xi+1\leq i\leq\chi,\\
        2+\beta&\text{ if }\,\,\chi+1\leq i\leq N,\\
    \end{cases}
    \label{ainew}
\end{equation}
with the parameters
\begin{equation}
           \alpha = \frac{l+h}{k},\,\,\,\,\,\,
        \beta = \frac{l}{k},\,\,\,\,\,\,
        \gamma = \frac{l+p}{k},
   \end{equation}
satisfying the condition $\beta<\gamma<\alpha$.
By introducing the matrix $\mathcal{A}$ and the vectors $\vec{y}$ and $\vec{v}$, we are able to write the total energy as
\begin{eqnarray}
\nonumber
    \Phi_H &=& \frac{k}{2}\vec{y}\cdot\mathcal{A}\vec{y} - kY\vec{v}\cdot\vec{y} + \frac{k}{2}Y^2 + \frac{l}{2}NY^2\\
    \label{enevec}
    &&+\frac{p}{2}Y_b^2(\eta-\chi)+(\eta-\xi)\Delta E.
\end{eqnarray}
We suppose to embed the system in a thermal bath at temperature $T$ and, assuming to be not far from the thermodynamic equilibrium, we can evaluate the partition function
\begin{equation}
    Z_H(Y)=\sum_{\chi=0}^\eta\sum_{\xi=0}^\chi\int_{\mathbb{R}^N} e^{-\frac{\Phi_H}{K_BT}}\mathrm{d} \vec{y}.
\end{equation}
By using Eq.(\ref{enevec}), it can be evaluated as
\begin{equation}
    \label{eq:fra:soft_part_fun}
    Z_H(Y)=\sum_{\chi=0}^\eta\sum_{\xi=0}^\chi \mathcal{I}_{\xi,\chi}e^{\lambda_{\xi,\chi}},
\end{equation}
where
\begin{equation}
    \lambda_{\xi,\chi} = -\frac{kY^2}{2K_BT}-\frac{lNY^2}{2K_BT}-\frac{\Delta E}{K_BT}(\eta-\xi)-\frac{pY_b^2}{2K_BT}(\eta-\chi),
\end{equation}
and
\begin{equation}
    \mathcal{I}_{\xi,\chi} = \int_{\mathbb{R}^N}\exp{\left( -\frac{k}{2K_BT}\vec{y}\cdot\mathcal{A}\vec{y}+\frac{kY}{K_BT}\vec{v}\cdot\vec{y}\right) }\mathrm{d} \vec{y}.
\end{equation}
 Using Eq.\eqref{eq:fra:Gauss_matrix} we get 
 \begin{equation}
    \mathcal{I}_{\xi,\chi} =\sqrt{\frac{(2\pi K_BT)^N}{k^N\det \mathcal{A}}}\exp{\left( \frac{kY^2}{2K_BT}\vec{v}\cdot\mathcal{A}^{-1}\vec{v}\right) }.
\end{equation}
Summing up, we obtain the partition function as
\begin{equation}
    \begin{aligned}
        Z_H(Y)=&\sum_{\chi=0}^\eta\sum_{\xi=0}^\chi\sqrt{\frac{(2\pi K_BT)^N}{k^N\det \mathcal{A}}} e^{-\frac{kY^2}{2K_BT}\left(1+\beta N-\vec{v}\cdot\mathcal{A}^{-1}\vec{v}\right)}\\
        &\times\exp{\left[ -\frac{\Delta E}{K_BT}(\eta-\xi)-\frac{pY_b^2}{2K_BT}(\eta-\chi)\right] }.
    \end{aligned}
\end{equation}
 In this case, by using Eq.(\ref{deri}), we get that the expected value of the applied force is
\begin{equation}
    \label{eq:fra:exact_soft_f}
    \langle f \rangle=\frac{\displaystyle{\sum_{\chi=0}^\eta\sum_{\xi=0}^\chi(\det\mathcal{A})^{-\frac{1}{2}}\exp{\left( \frac{2\Delta E\xi+pY_b^2\chi-kY^2q}{2K_BT}\right) }q}}{\displaystyle{\sum_{\chi=0}^\eta\sum_{\xi=0}^\chi(\det\mathcal{A})^{-\frac{1}{2}}\exp{\left( \frac{2\Delta E\xi+pY_b^2\chi-kY^2q}{2K_BT}\right) }}}k\,Y,
\end{equation}
where we used the definition of $q$ in Eq.\eqref{eq:fra:q}.
This is the expression for the average value of the force necessary to impose the extension $Y$ to the system. 
Similarly, we obtain the interfaces positions
\begin{align}
\label{xifinex}
    \langle \xi \rangle=& \frac{\displaystyle{\sum_{\chi=0}^\eta\sum_{\xi=0}^\chi(\det\mathcal{A})^{-\frac{1}{2}}\exp{\left( \frac{2\Delta E\xi+pY_b^2\chi-kY^2q}{2K_BT}\right) }\xi}}{\displaystyle{\sum_{\chi=0}^\eta\sum_{\xi=0}^\chi(\det\mathcal{A})^{-\frac{1}{2}}\exp{\left( \frac{2\Delta E\xi+pY_b^2\chi-kY^2q}{2K_BT}\right) }}},\\
    \label{chifinex}
    \langle \chi \rangle=& \frac{\displaystyle{\sum_{\chi=0}^\eta\sum_{\xi=0}^\chi(\det\mathcal{A})^{-\frac{1}{2}}\exp{\left( \frac{2\Delta E\xi+pY_b^2\chi-kY^2q}{2K_BT}\right) }\chi}}{\displaystyle{\sum_{\chi=0}^\eta\sum_{\xi=0}^\chi(\det\mathcal{A})^{-\frac{1}{2}}\exp{\left( \frac{2\Delta E\xi+pY_b^2\chi-kY^2q}{2K_BT}\right) }}}.
\end{align}
These results allow us to fully describe the fracture behavior for a ductile material. In particular, based on Eqs.(\ref{xifinex}) and (\ref{chifinex}), we are able to determine when the fracture is brittle, without the region of softened elements, or when the fracture is ductile, i.e., with a non negligible fraction of softened elements, representing the cohesive region.

We mentioned the Dugdale and Barenblatt models since historically they are the most important approaches to introduce a process zone in fracture phenomena.  It is useful to remember that the original Dugdale model has been developed for plane strain conditions. Other approaches have been developed successively to consider plane stress conditions \cite{stress1,stress2}. 
However, our model is composed of a quasi one-dimensional lattice of springs that does not allow the access to realistic elastic fields in the structure.  Hence, it is difficult to quantitatively compare our results with elastic models in both plane stress and plane strain. Moreover, Dugdale model does not account for hardening phenomena, such as our approach, which is completely linear. In spite of these limitations, Dugdale model has been generalized for strain hardening materials \cite{hard1,hard2}. It is also important to remember that the process zone in real situations extends beyond the fracture growth plane, a point neglected in both Dugdale's original model and ours. In real fractures, the actual deformation is represented by a complicated three-dimensional field, completely disregarded in our one-dimensional analysis. To conclude, the purpose of our models is not to improve aspects related to continuous elastic fields but rather to introduce the effects of temperature into a simplified model. With this in mind, our approaches are not created to replace classical ones but only to inform them of how temperature acts in fracture phenomena.

It is also important to discuss the physical meaning of the softened state of breakable springs. The ductility in metallic materials is related to a population of dislocations originated by the moving crack, generating a damaged zone near the crack tip with degraded elastic properties \cite{dislo1,dislo2,dislo3,dislo4}. Since, we do not have the possibility to consider realistic dislocations in our model, we introduced the weakened state for the breakable springs, corresponding to the degraded elastic properties of the damaged zone. In the realistic case, the brittle-to-ductile transition is controlled by the competition between continuing the fracture (as in the brittle case) or using an amount of energy to generate dislocations that degrade the material. In our model, we have similar competition between the intact/broken switching (brittle regime) or the intact/softened switching (ductile regime). This competition is strongly influenced by temperature and our model explains this effect in detail. Besides metals, a similar damaged zone, describing the physical state of the material between the intact and the fully broken conditions, has been also observed in different systems including concrete \cite{huwi}, soft materials \cite{creton,trentadue}, polymeric networks \cite{awa,buche}, bones \cite{kir}.

\section{Asymptotic behavior of the softening-fracture model}
\label{asysoft}

 Once again to obtain clearer analytic results, we consider the behavior of systems with large values of $N$.
We have (see Appendix \ref{appendixb})
\begin{equation}    
\label{qsofttext}
        q\sim \beta^2\left(\frac{\xi}{\gamma}-\frac{\xi}{\alpha}-\frac{\chi}{\gamma}+\frac{\chi}{\beta}\right)+\epsilon,
    \end{equation}
where $\epsilon$ is given in Eq.(\ref{eptext}), and
\begin{equation}
\label{detsofttext}
    \frac{\ln\det\mathcal{A}}{N} \sim \ln\tau_\beta+\frac{\xi}{N}\ln\frac{\tau_\alpha}{\tau_\gamma}+\frac{\chi}{N}\ln\frac{\tau_\gamma}{\tau_\beta},
\end{equation}
where $\tau_\alpha, \tau_\beta$ and $\tau_\gamma$ are given in Eq.(\ref{tautau}), for $s=\alpha,\beta,\gamma$.
Thus, we can write the expressions for the average force and the average interface positions as it follows
\begin{gather}
    \langle f \rangle\sim \frac{\sum_{\chi=0}^\eta\sum_{\xi=0}^\chi\left[\beta^2\left(\frac{\xi}{\gamma}-\frac{\xi}{\alpha}-\frac{\chi}{\gamma}+\frac{\chi}{\beta}\right)+\epsilon\right]e^{q_\xi\xi+q_\chi\chi}}{\sum_{\chi=0}^\eta\sum_{\xi=0}^\chi e^{q_\xi\xi+q_\chi\chi}}k\, Y,\\
    \label{xisoft}
    \langle \xi \rangle\sim \frac{\sum_{\chi=0}^\eta\sum_{\xi=0}^\chi\xi e^{q_\xi\xi+q_\chi\chi}}{\sum_{\chi=0}^\eta\sum_{\xi=0}^\chi e^{q_\xi\xi+q_\chi\chi}},\\
    \label{chisoft}
    \langle \chi \rangle\sim \frac{\sum_{\chi=0}^\eta\sum_{\xi=0}^\chi\chi e^{q_\xi\xi+q_\chi\chi}}{\sum_{\chi=0}^\eta\sum_{\xi=0}^\chi e^{q_\xi\xi+q_\chi\chi}},
\end{gather}
where we introduced the  quantities
\begin{align}
    q_\xi&=-\frac{1}{2}\ln\frac{\tau_\alpha}{\tau_\gamma}-\frac{kY^2\beta^2}{2K_BT}\left(\frac{1}{\gamma}-\frac{1}{\alpha}\right)+\frac{\Delta E}{K_BT},\\
    q_\chi&=-\frac{1}{2}\ln\frac{\tau_\gamma}{\tau_\beta}-\frac{kY^2\beta^2}{2K_BT}\left(\frac{1}{\beta}-\frac{1}{\gamma}\right)+\frac{pY_b^2}{2K_BT}.
\end{align}
Using the expressions for $\langle\xi\rangle$ and $\langle\chi\rangle$, we can rewrite $\langle f\rangle$ in the simpler form
\begin{equation}
    \langle f \rangle\sim \left(\epsilon +\beta^2\left(\frac{1}{\gamma}-\frac{1}{\alpha}\right)\langle \xi \rangle+\beta^2\left(\frac{1}{\beta}-\frac{1}{\gamma}\right)\langle \chi \rangle \right ) k\, Y.
    \label{forzafin}
\end{equation}
Hence, once we know the expected values $\langle\xi\rangle$ and $\langle\chi\rangle$ of the interfaces positions,  we also know the force required to impose the extension $Y$. 
To simplify the notation, we introduce 
\begin{align}
    w = e^{q_\xi},\\
    z = e^{q_\chi}.
\end{align}
By Eqs.(\ref{xisoft}) and (\ref{chisoft}), using   Eqs.\eqref{eq:fra:sums_1} and \eqref{eq:fra:sums_2},after long but straightforward calculations, we obtain 
\begin{align}
    \label{xifin}
    \langle \xi \rangle&=\frac{N_\xi}{D},\\
    \label{chifin}
    \langle \chi \rangle&=\frac{N_\chi}{D},
\end{align}
where, for the sake of readability, we introduced
\begin{align}
    &\begin{aligned}
        N_\xi =&\frac{w}{1-w}\bigg\{\frac{1-z^{\eta+1}}{1-z}-\frac{1-(wz)^{\eta+1}}{1-wz}\\
        &-\frac{(1-w)wz}{(1-wz)^2}\left[1-(wz)^\eta(1+\eta)+\eta(wz)^{\eta+1}\right]\bigg\},
    \end{aligned}\\
    &\begin{aligned}
        N_\chi =&\frac{z}{(1-z)^2}\left(1-z^\eta(1+\eta)+\eta z^{\eta+1}\right)\\
        &-\frac{w^2z}{(1-wz)^2}\left[1-(wz)^\eta(1+\eta)+\eta(wz)^{\eta+1}\right],
    \end{aligned}\\
    &\begin{aligned}
    D=&\frac{1-z^{\eta+1}}{1-z}-w\frac{1-(wz)^{\eta+1}}{1-wz}.
    \end{aligned}
\end{align}
These results approximate the behavior of the fracture process in the presence of the softening phenomenon for large values of $N$.
In particular, we can determine the limit for $N\to\infty$ of the main observables, in order to provide a precise physical interpretation of the brittle-to-ductile transition.
The obtained expressions depend on $w^\eta$, $z^\eta$ and $(wz)^\eta$ and, since in our model $\eta=N(1-\phi)$ where $\phi$ is the fraction of initially absent breakable springs, they present an exponent going to infinity when $N\to\infty$.
We know that, when $N\to\infty$, a generic power $x^N$ tends to infinity if $x>1$ and tends to zero if $\vert x \vert<1$, hence we study the three inequalities $w>1$, $z>1$ and $wz>1$, which will be useful to better understand the system behavior.
These inequalities are equivalent to study the positive character of their exponents $q_\xi$, $q_\chi$, and $q_\xi+q_\chi$.

\begin{figure}
    \centering
    \includegraphics[scale=1]{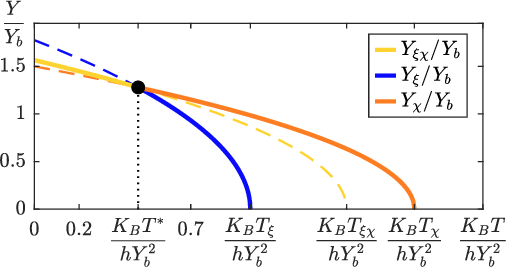}
    \caption{Behavior of the dimensionless extension thresholds $Y_\xi/Y_b$, $Y_\chi/Y_b$ and $Y_{\xi\chi}/Y_b$  versus the dimensionless temperature $K_BT/(hY_b^2)$ of the system. From the physical point of view, $Y_{\xi\chi}$ (yellow or light gray curve) describes the brittle fracture below the transition temperature $T^*$, and the couple $Y_\xi$ (blue or dark gray), $Y_\chi$ (orange or intermediate gray) describes the ductile fracture above the temperature $T^*$. The three curves $Y_\xi/Y_b$, $Y_\chi/Y_b$ and $Y_{\xi\chi}/Y_b$  versus  $K_BT/(hY_b^2)$ intersect at the bifurcation black point, characterized by $T^*$. We adopted the parameters $\alpha=7/5$, $\beta=2/5$, $\gamma=9/10$, and $\Delta E/(hY_b^2)=1/10$.}
    \label{fig:FRA:InfN_Critic_Softened}
\end{figure}

\begin{figure*}[t!]
    \centering
    \includegraphics[scale=1]{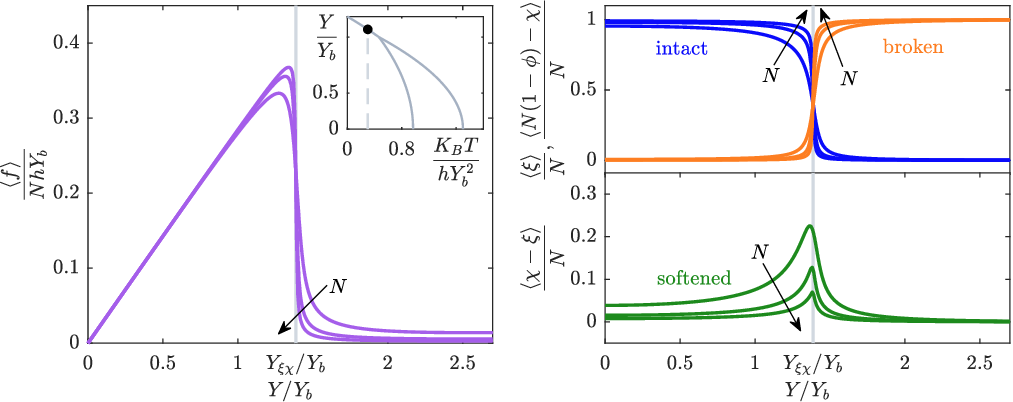}
    \caption{Brittle response of the fracture phenomenon ($0<T<T^*$). Left panel: dimensionless force versus dimensionless extension for different values of $N=100,\,250,\,500$ (as indicated by the arrow). Right panel: fraction of intact, softened and broken elements for different values of $N=100,\,250,\,500$ (as indicated by arrows). Inset: same plot as Fig.\ref{fig:FRA:InfN_Critic_Softened}, where the temperature used here is indicated by the black point. We adopted the parameters $\alpha=7/5$, $\beta=2/5$, $\gamma=9/10$, $\Delta E/(hY_b^2)=1/10$, $\phi=0$ (i.e., $\eta=N$), and $K_B T/(hY_b^2)=3/10$.}
    \label{fig:FRA:LargeN_Fracture_Soft_Pip_BRIT}
\end{figure*}

We start by setting the exponent of $w$ larger than zero
\begin{equation}
    q_\xi=-\frac{1}{2}\ln\frac{\tau_\alpha}{\tau_\gamma}-\frac{kY^2\beta^2}{2K_BT}\left(\frac{1}{\gamma}-\frac{1}{\alpha}\right)+\frac{\Delta E}{K_BT}>0.
\end{equation}
In terms of  $Y$, this inequality gives 
\begin{equation}
    \lvert Y \rvert <\sqrt{\frac{1}{k\beta^2\left(\frac{1}{\gamma}-\frac{1}{\alpha}\right)}\left[(h-p)Y_p^2-K_BT\ln\frac{\tau_\alpha}{\tau_\gamma}\right]}\triangleq Y_\xi,
\end{equation}
where we introduced a first extension threshold $Y_\xi$.
We observe that $k\beta^2\left(\frac{1}{\gamma}-\frac{1}{\alpha}\right)$ is always positive because $p<h$ by definition and, then, the argument of the square root is positive when the temperature $T$ is smaller than the critical temperature $T_\xi$ defined as
\begin{equation}
    T_\xi=\frac{(h-p)Y_p^2}{K_B\ln\frac{\tau_\alpha}{\tau_\gamma}}.
\end{equation}
The meaning of $Y_\xi$ and $T_\xi$ will be clarified later. 
By setting the exponent of $z$ greater than zero, we define the inequality
\begin{equation}
    q_\chi=-\frac{1}{2}\ln\frac{\tau_\gamma}{\tau_\beta}-\frac{kY^2\beta^2}{2K_BT}\left(\frac{1}{\beta}-\frac{1}{\gamma}\right)+\frac{pY_b^2}{2K_BT}>0.
\end{equation}
It can be solved with respect to $Y$, eventually giving the result
\begin{equation}
    \lvert Y \rvert <\sqrt{\frac{1}{k\beta^2\left(\frac{1}{\beta}-\frac{1}{\gamma}\right)}\left[pY_b^2-K_BT\ln\frac{\tau_\gamma}{\tau_\beta}\right]}\triangleq Y_\chi,
\end{equation}
where we introduced a second extension threshold $Y_\chi$.
As before, the quantity $k\beta^2\left(\frac{1}{\beta}-\frac{1}{\gamma}\right)$ is always positive and therefore the whole square root argument is positive for values of the temperature below the critical temperature $T_\chi$ defined as
\begin{equation}
    T_\chi=\frac{pY_b^2}{K_B\ln\frac{\tau_\gamma}{\tau_\beta}}.
\end{equation}
As before, $Y_\chi$ and $T_\chi$ will be physically interpreted in the following.  
Finally, we set the exponent of $wz$ greater than zero, which corresponds to $q_\xi+q_\chi>0$. We obtain the inequality 
\begin{equation}
    \lvert Y \rvert <\sqrt{\frac{1}{k\beta^2\left(\frac{1}{\beta}-\frac{1}{\alpha}\right)}\left[pY_b^2+(h-p)Y_p^2-K_BT\ln\frac{\tau_\alpha}{\tau_\beta}\right]}\triangleq Y_{\xi\chi},
\end{equation}
where we introduced a third extension threshold $Y_{\xi\chi}$. Being $k\beta^2\left(\frac{1}{\beta}-\frac{1}{\alpha}\right)$ always positive, the square root has a  positive argument when $T<T_{\xi\chi}$, where
\begin{equation}
    T_{\xi\chi}=\frac{pY_b^2+(h-p)Y_p^2}{K_B\ln\frac{\tau_\alpha}{\tau_\beta}}. 
\end{equation}
Again, we will discuss in the following the physical meaning of $Y_{\xi\chi}$ and $T_{\xi\chi}$. 
Summarizing these results, we can write
\begin{align}
    &\begin{cases}
    w>1\\
    q_\xi>0
    \end{cases}
    \Leftrightarrow
    \lvert Y \rvert <\sqrt{\frac{(l+p)(l+h)Y_p^2}{l^2}\left(1-\frac{T}{T_\xi}\right)},\\
    &\begin{cases}
    z>1\\
    q_\chi>0
    \end{cases}
    \Leftrightarrow
    \lvert Y \rvert <\sqrt{\frac{(l+p)Y_b^2}{l}\left(1-\frac{T}{T_\chi}\right)},\\
    &\begin{cases}
    wz>1\\
    q_\xi+q_\chi>0
    \end{cases}
    \Leftrightarrow
    \lvert Y \rvert <\sqrt{\frac{pY_b^2+(h-p)Y_p^2}{\frac{lh}{l+h}}\left(1-\frac{T}{T_{\xi\chi}}\right)},
\end{align}
where we used the definition for the critical temperatures $T_{\xi}$, $T_{\chi}$ and $T_{\xi\chi}$ previously introduced.

\begin{figure*}[t!]
    \centering
    \includegraphics[scale=1]{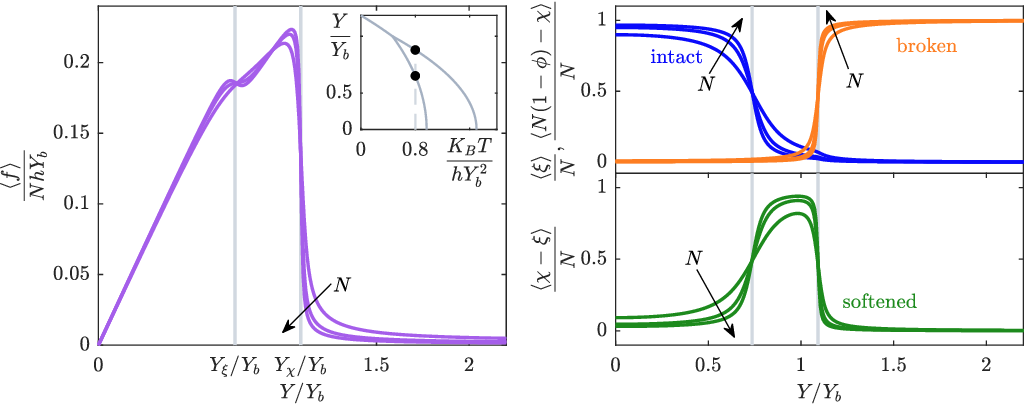}
    \caption{Ductile response of the fracture phenomenon ($T^*<T<T_\xi$). Left panel: dimensionless force versus dimensionless extension for different values of $N=500,\,1000,\,1500$ (as indicated by the arrow). Right panel: fraction of intact, softened and broken elements for different values of $N=500,\,1000,\,1500$ (as indicated by arrows). Inset: same plot as Fig.\ref{fig:FRA:InfN_Critic_Softened}, where the temperature used here is indicated by the black points. We adopted the parameters $\alpha=7/5$, $\beta=2/5$, $\gamma=9/10$, $\Delta E/(hY_b^2)=1/10$, $\phi=0$ (i.e., $\eta=N$), and $K_BT/(hY_b^2)=4/5$.}
    \label{fig:FRA:LargeN_Fracture_Soft_Pip_DUCT}
\end{figure*}

The three dimensionless extension thresholds $Y_\xi/Y_b$, $Y_\chi/Y_b$ and $Y_{\xi\chi}/Y_b$ are plotted versus the dimensionless temperature $K_BT/(hY_b^2)$ in Fig.\ref{fig:FRA:InfN_Critic_Softened}. We can already anticipate that brittle or ductile behavior depends on the sign of  $Y_\chi-Y_\xi$. Indeed, we have a brittle fracture if $Y_\chi<Y_\xi$, and a ductile fracture if  $Y_\chi>Y_\xi$ (see Fig.\ref{fig:FRA:InfN_Critic_Softened}). In the first brittle case, the rupture occurs for $Y=Y_{\xi\chi}$. In the second ductile case, the softening occurs for $Y=Y_\xi$ and rupture for $Y=Y_\chi$.
We observe therefore that it exists a brittle-to-ductile transition temperature $T^*$ that separates the brittle behavior from  the ductile one (see the yellow point in Fig.\ref{fig:FRA:InfN_Critic_Softened}). This temperature is defined by equating $Y_\xi$ and $Y_\chi$, as follows
\begin{equation}
    \sqrt{\frac{(h-p)Y_p^2}{k\beta^2\left(\frac{1}{\gamma}-\frac{1}{\alpha}\right)}\left(1-\frac{T^*}{T_\xi}\right)}=\sqrt{\frac{pY_b^2}{k\beta^2\left(\frac{1}{\beta}-\frac{1}{\gamma}\right)}\left(1-\frac{T^*}{T_\chi}\right)}.
\end{equation}
When solved, this equation gives the value of $T^*$ as
\begin{equation}
    \frac{K_BT^*}{hY_b^2} = \frac{\left(\frac{1}{\alpha-\beta}\right)\left[\alpha\left(\frac{Y_p}{Y_b}\right)^2-\beta\right]}{\left(\frac{\alpha}{\alpha-\gamma}\right)\ln\frac{\tau_\alpha}{\tau_\gamma}-\left(\frac{\beta}{\gamma-\beta}\right)\ln\frac{\tau_\gamma}{\tau_\beta}}.
\end{equation}
Interestingly, this quantity can be also explicitly written in terms of the elastic constants of the system
\begin{equation}
    T^* = \frac
    {(l+h)Y_p^2-lY_b^2}
    {K_B\left(\frac{l+h}{h-p}\ln\frac{\tau_\alpha}{\tau_\gamma}-\frac{l}{p}\ln\frac{\tau_\gamma}{\tau_\beta}\right)}.
    \label{tempaste}
\end{equation}

\begin{figure*}[t!]
    \centering
    \includegraphics[scale=1]{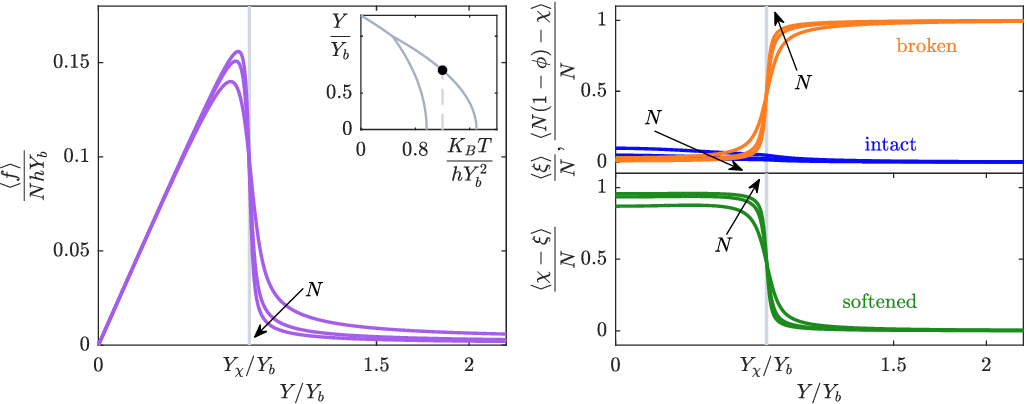}
    \caption{Over-ductile response of the fracture phenomenon ($T_\xi<T<T_\chi$). Left panel: dimensionless force versus dimensionless extension for different values of $N=500,\,1000,\,1500$ (as indicated by the arrow). Right panel: fraction of intact, softened and broken elements for different values of $N=500,\,1000,\,1500$ (as indicated by arrows). Inset: same plot as Fig.\ref{fig:FRA:InfN_Critic_Softened}, where the temperature used here is indicated by the black point. We adopted the parameters $\alpha=7/5$, $\beta=2/5$, $\gamma=9/10$, $\Delta E/(hY_b^2)=1/10$, $\phi=0$ (i.e., $\eta=N$), and $K_BT/(hY_b^2)=6/5$.}
    \label{fig:FRA:LargeN_Fracture_Soft_Pip_OVERDUCT}
\end{figure*}

To justify the introduction of all these quantities and notations, we use now Eqs.(\ref{forzafin}), (\ref{xifin}) and (\ref{chifin}) to observe the behavior of the system with different values of $N$ and temperature $T$. 
In particular, we consider Fig.\ref{fig:FRA:InfN_Critic_Softened} and we show the system behavior for three values of the temperature belonging to the regions $0<T<T^*$ (brittle response, Fig.\ref{fig:FRA:LargeN_Fracture_Soft_Pip_BRIT}), $T^*<T<T_\xi$ (ductile response, Fig.\ref{fig:FRA:LargeN_Fracture_Soft_Pip_DUCT}), and  $T_\xi<T<T_\chi$ (over-ductile response, Fig.\ref{fig:FRA:LargeN_Fracture_Soft_Pip_OVERDUCT}). 
We do not consider values of the temperature larger than $T_\chi$ since, in this case, all elements are broken  due to the only thermal effects, without the application of mechanical actions.

In Fig.\ref{fig:FRA:LargeN_Fracture_Soft_Pip_BRIT}, we can find the dimensionless force given by Eq.(\ref{forzafin}) in the first panel, and the three quantities $\langle \xi \rangle/N$, $\langle \chi-\xi \rangle/N$ and $\langle N(1-\phi)-\chi \rangle/N$ representing the fraction of intact, softened and broken elements, calculated through Eqs.(\ref{xifin}) and (\ref{chifin}), respectively, in the second panel. 
For simplicity, we always considered $\phi=0$. Note that the dimensionless force is divided by $N$ so as to be consistent with the definition of mechanical stress.
We can see that, with a temperature in the range $0<T<T^*$, the force drops to zero in correspondence to the extension threshold $Y_{\xi\chi}$, describing the simultaneous rupture of all elements. 
Indeed, it can be seen in the second panel that the elements change almost completely from the intact to the broken state, with a fraction of softened elements that is negligible. 
The response becomes increasingly sharp as the value of $N$ increases. In particular, the fraction of softened elements decreases to zero for $N$ growing. A direct transition from intact to broken elements without an intermediate phase is therefore observed.
This confirms that the response is brittle for $T<T^*$ and the rupture of the system occurs in this case at the applied extension $Y_{\xi\chi}$.

In Fig.\ref{fig:FRA:LargeN_Fracture_Soft_Pip_DUCT}, we represent the same functions for a temperature in the range $T^*<T<T_\xi$. 
In this case, we observe an almost simultaneous transition of all breakable elements from the intact to the softened state at the extension $Y_\xi$ and a subsequent transition from the softened to the broken state at the threshold $Y_\chi$. 
This behavior reproduces a ductile fracture, and the intermediate phase, characterized by the softened elements, mimics the cohesive phase of the fracture phenomenon.
Also, in the force-extension diagram we see a first peak in correspondence to the softening of the elements, and a second peak describing the actual rupture. This curve is sharper for high values of $N$ and smoother for low values. 
The comparison of Figs.\ref{fig:FRA:LargeN_Fracture_Soft_Pip_BRIT} and \ref{fig:FRA:LargeN_Fracture_Soft_Pip_DUCT} shows the transition from a brittle to a ductile fracture as temperature increases, $T^*$ being the threshold temperature between the two regimes. 
This transition is described by the bifurcation at $T^*$ exhibited in Fig.\ref{fig:FRA:InfN_Critic_Softened} (see the yellow point), which gives rise to the intermediate region with softened elements.

To complete the picture on the system behavior, we also show in Fig.\ref{fig:FRA:LargeN_Fracture_Soft_Pip_OVERDUCT} the extreme situation when temperature is in the range $T_\xi<T<T_\chi$ (over-ductile response).  Since the temperature is larger than $T_\xi$, at the beginning of the traction almost all elements are  already in the softened state even without an applied mechanical action. 
 As a result, we can observe only one transition between the softened state  and the broken state at the extension threshold $Y_\chi$. Consequently, in this temperature range, we observe a brittle transition between thermally-softened and broken elements. 
Since this response is observed only after the classical ductile behavior, we called it over-ductile response.

\begin{figure}
    \centering
    \includegraphics[scale=1]{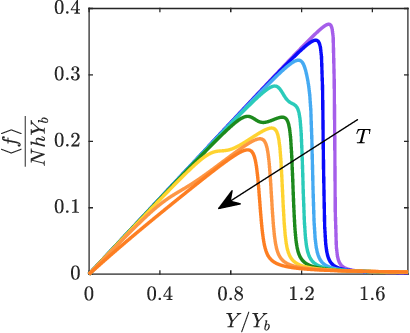}
    \caption{Dimensionless force versus dimensionless extension for different values of the thermal to elastic energy ratio $K_BT/(hY_b^2)=\{0.3,0.4,0.5,0.6,0.7,0.8,0.9,1\}$ (as indicated by the arrow). We observe that the behavior of the model changes from brittle, at low values of temperature, ductile for intermediate temperatures, to over ductile at high temperatures. We adopted the parameters $N=1000$, $\alpha=7/5$, $\beta=2/5$, $\gamma=9/10$, $\phi=0$, and $\Delta E/(hY_b^2)=1/10$.}
    \label{fig:fra:FTransition}
\end{figure}

To better visualize the transitions between the different fracture regimes, we show in Fig.\ref{fig:fra:FTransition} some force-extension curves corresponding to different temperatures, spanning over brittle, ductile, and over-ductile regimes. 
It is interesting to remark that, within the ductile fracture, the shape of the force-extension curve is smoother in correspondence to the system softening/breaking since the cohesive phase is able to absorb an amount of energy before the final rupture. 
We also note that, within the ductile regime, the softening stress is higher than the failure stress for lower temperatures and conversely the softening stress becomes lower than the failure stress for higher temperatures. This point will be further discussed below.
It is  important to underline that all curves seen in Figs.\ref{fig:FRA:LargeN_Fracture_Soft_Pip_BRIT}, \ref{fig:FRA:LargeN_Fracture_Soft_Pip_DUCT},  \ref{fig:FRA:LargeN_Fracture_Soft_Pip_OVERDUCT}, and \ref{fig:fra:FTransition} have been obtained through Eqs.(\ref{forzafin}), (\ref{xifin}) and (\ref{chifin}) with a large, but finite value of $N$. In the following we also describe the thermodynamic limit $N\to\infty$.

\begin{figure*}
    \centering
    \includegraphics[scale=1]{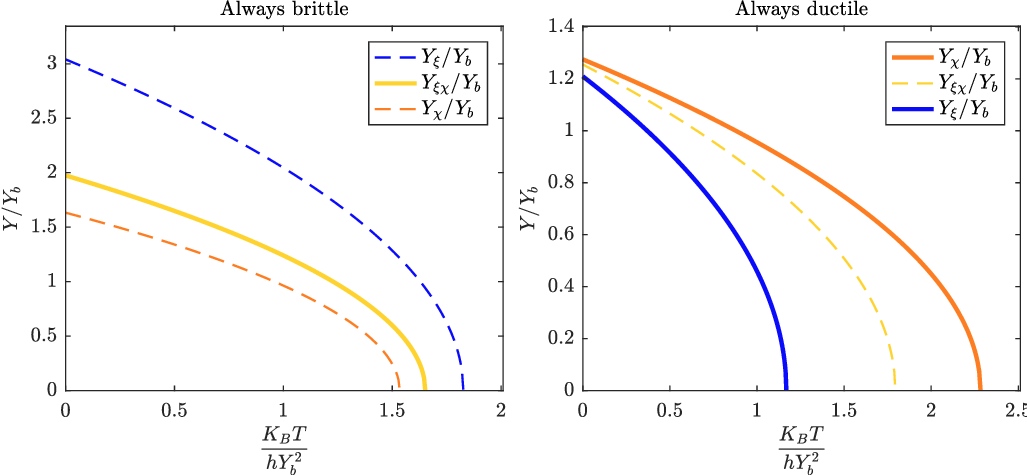}
    \caption{Behavior of the dimensionless extension thresholds $Y_\xi/Y_b$, $Y_\chi/Y_b$ and $Y_{\xi\chi}/Y_b$  versus the dimensionless temperature $K_BT/(hY_b^2)$ for a system always brittle (left panel), and for a system always ductile (right panel). In the left panel we have $Y_\chi<Y_{\xi\chi}<Y_\xi$ and we adopted the parameters $\alpha=13/10$, $\beta=3/10$, $\gamma=4/5$, and $\Delta E/(hY_b^2)=1/5$. In the right panel we have $Y_\chi>Y_{\xi\chi}>Y_\xi$ and we adopted the parameters $\alpha=9/5$, $\beta=4/5$, $\gamma=13/10$, and $\Delta E/(hY_b^2)=1/10$.}
    \label{inhe-bri-duc}
\end{figure*}

What has been described so far represents the modeling of brittle-to-ductile transition induced by thermal fluctuations. Our model also allows us to describe a parametric brittle-to-ductile transition, i.e., intrinsic to the structure of the system. 
This means that there can be systems that exhibit only brittle or ductile behavior, regardless of temperature. 
On the one hand, an example of always brittle system is shown in Fig.\ref{inhe-bri-duc}, left panel, where we represent the three dimensionless extension thresholds $Y_\xi/Y_b$, $Y_\chi/Y_b$ and $Y_{\xi\chi}/Y_b$ versus the dimensionless temperature $K_BT/(hY_b^2)$. We can see that $Y_\chi<Y_{\xi\chi}<Y_\xi$ for any value of the temperature. This means that there is no temperature high enough to induce a ductile fracture ($T^*>T_\xi$). 
On the other hand,  an example of always ductile system is shown in Fig.\ref{inhe-bri-duc}, right panel. In this case,  $Y_\chi>Y_{\xi\chi}>Y_\xi$ for any value of the temperature  so that there is no temperature low enough to  induce a brittle fracture ($T^*<0$).
These two situations describe materials that are always brittle or always ductile, regardless of the considered temperature. 

To conclude, we obtained two types of brittle-to-ductile transitions: a \textit{thermal} transition, induced by the effects of thermal fluctuations, and a \textit{parametric} transition, induced by the values of the elastic parameters.  
It is worth to point out that this rather rich fracture behavior has been obtained based on a minimal system depending on the competition between elastic, entropic and fracture energy terms, regulated by the temperature and material parameters.

\section{Thermodynamic limit of the softening-fracture model}
\label{thermosoft}

 To give an even clearer physical description, we deduce  here analytic results in the thermodynamic limit, $N\to\infty$.
We start the analysis by examining the average value of the number of intact, softened and broken elements of the system.
If we consider the brittle behavior, with $0<T<T^*$, the thermodynamic limit gives
\begin{eqnarray}
&&\lim_{N\to\infty}\frac{\langle \xi \rangle}{N}=\left \{ \begin{array}{cc}
   1-\phi  & \mbox{ if } Y<Y_{\xi\chi}, \\
   0  & \mbox{ if } Y>Y_{\xi\chi},
\end{array}\right.\\
&&\lim_{N\to\infty}\frac{\langle \chi-\xi \rangle}{N}=0 \mbox{  for all } Y,\\
&&\lim_{N\to\infty}\frac{\langle N(1-\phi)-\chi \rangle}{N}=\left \{ \begin{array}{cc}
    0 & \mbox{ if } Y<Y_{\xi\chi}, \\
   1-\phi  & \mbox{ if } Y>Y_{\xi\chi}.
\end{array}\right.
\end{eqnarray}
In this case, we observe a direct transition between intact and broken elements without going through the softened state.
However, if we take into account the ductile behavior with $T^*<T<T_\xi$, we obtain for $N\to\infty$
\begin{eqnarray}
&&\lim_{N\to\infty}\frac{\langle \xi \rangle}{N}=\left \{ \begin{array}{cc}
   1-\phi  & \mbox{ if } Y<Y_{\xi}, \\
   0  & \mbox{ if } Y>Y_{\xi},
\end{array}\right.\\
&&\lim_{N\to\infty}\frac{\langle \chi-\xi \rangle}{N}=\left \{ \begin{array}{cl} 
      0   & \mbox{ if } Y<Y_{\xi}, \\
   1-\phi  & \mbox{ if } Y_{\xi}<Y<Y_{\chi}, \\
   0  & \mbox{ if } Y>Y_{\chi},
\end{array}\right.\\
&&\lim_{N\to\infty}\frac{\langle N(1-\phi)-\chi \rangle}{N}=\left \{ \begin{array}{cc}
    0 & \mbox{ if } Y<Y_{\chi}, \\
   1-\phi  & \mbox{ if } Y>Y_{\chi}.
\end{array}\right.
\end{eqnarray}
 In this case, we observe the emergence of a region with softened elements, corresponding to the cohesive zone.
Finally, the over-ductile regime, characterized by $T_\xi<T<T_\chi$, for $N\to\infty$ leads to
\begin{eqnarray}
&&\lim_{N\to\infty}\frac{\langle \xi \rangle}{N}= 0 \mbox{  for all } Y,\\
&&\lim_{N\to\infty}\frac{\langle \chi-\xi \rangle}{N}=\left \{ \begin{array}{cc}
    1-\phi & \mbox{ if } Y<Y_{\chi}, \\
    0 & \mbox{ if } Y>Y_{\chi},
\end{array}\right.\\
&&\lim_{N\to\infty}\frac{\langle N(1-\phi)-\chi \rangle}{N}=\left \{ \begin{array}{cc}
    0 & \mbox{ if } Y<Y_{\chi}, \\
   1-\phi  & \mbox{ if } Y>Y_{\chi}.
\end{array}\right.
\end{eqnarray}
In this regime, all elements are initially softened and therefore the single transition corresponds to their complete breaking.

Concerning the expected value of the force, from Eq.(\ref{forzafin}), we can write
\begin{equation}
    \lim_{N\to\infty}\frac{\langle f \rangle}{N}=\left(\beta^2\left(\frac{1}{\gamma}-\frac{1}{\alpha} \right)\frac{\langle \xi \rangle}{N}+\beta^2\left(\frac{1}{\beta}-\frac{1}{\gamma} \right)\frac{\langle \chi \rangle}{N}\right) k\, Y,
    \label{forcelim}
\end{equation}
where we can substitute the values of $\langle \xi \rangle/N$ and $\langle \chi \rangle/N$ pertinent to each fracture regime. 
We remark that in Eq.(\ref{forcelim}), we have canceled out the first term shown in Eq.(\ref{forzafin}) since $N\to\infty$.
For the brittle behavior ($0<T<T^*$), we have
\begin{eqnarray}
    \lim_{N\to\infty}\frac{\langle f \rangle}{N}=\left \{ \begin{array}{cc}
   \beta^2(1-\phi)\left(\frac{1}{\beta}-\frac{1}{\alpha} \right) k\, Y& \mbox{ if } Y<Y_{\xi\chi}, \\
   0  & \mbox{ if } Y>Y_{\xi\chi}.
\end{array}\right.\label{brittlestress}
\end{eqnarray}

\begin{figure*}
    \centering
    \includegraphics[scale=1]{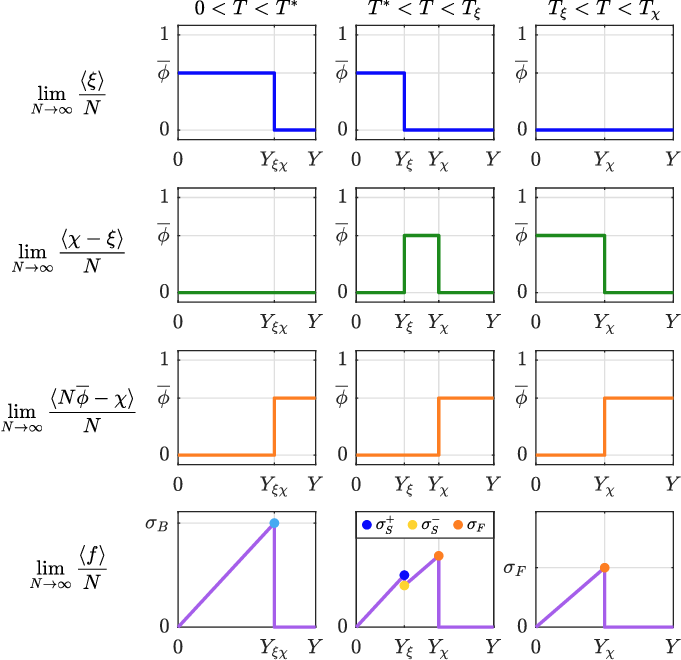}
    \caption{Response of the system in the thermodynamic limit within the three temperature regimes $0<T<T^*$ (brittle, first column), $T^*<T<T_\xi$ (ductile, second column), and $T_\xi<T<T_\chi$ (over-ductile, third column). We plotted the average number of intact (first row), softened (second row) and broken elements (third row), and the stress $\langle f \rangle/N$ (fourth row), for $N\to\infty$. To compact the notation, we defined $\Bar{\phi}=1-\phi$, and we introduced the characteristic stresses $\sigma_B$, $\sigma_S^+$, $\sigma_S^-$, and $\sigma_F$, as defined in Eqs.(\ref{stressB}), (\ref{stressS+}), (\ref{stressS-}) and (\ref{stressF}).}
    \label{fig:fra:infN_soft}
\end{figure*}

For the ductile behavior ($T^*<T<T_\xi$), we have
\begin{eqnarray}
\lim_{N\to\infty}\frac{\langle f \rangle}{N}=\left \{ \begin{array}{cl} 
      \beta^2(1-\phi)\left(\frac{1}{\beta}-\frac{1}{\alpha} \right) k\,Y  & \mbox{ if } Y<Y_{\xi}, \\
  \beta^2(1-\phi)\left(\frac{1}{\beta}-\frac{1}{\gamma} \right) k\,Y  & \mbox{ if } Y_{\xi}<Y<Y_{\chi},\,\,\,\,\,\, \\
   0  & \mbox{ if } Y>Y_{\chi}.
\end{array}\right.\label{ductilestress}
\end{eqnarray}
Finally, for the over-ductile behavior ($T_\xi<T<T_\chi$), we get
\begin{eqnarray}
    \lim_{N\to\infty}\frac{\langle f \rangle}{N}=\left \{ \begin{array}{cc}
  \beta^2(1-\phi)\left(\frac{1}{\beta}-\frac{1}{\gamma} \right)  k\,Y & \mbox{ if } Y<Y_{\chi}, \\
   0  & \mbox{ if } Y>Y_{\chi}.
\end{array}\right.
\end{eqnarray}
The thermodynamic limit behavior ($N\to\infty$) of the intact, softened and broken elements together with the  value of the stress $\langle f \rangle/N$ is exhibited in Fig.\ref{fig:fra:infN_soft}, where all the three fracture regimes brittle, ductile and over-ductile are considered. 
We remark that the resulting overall picture is coherent  with the plots in Figs.\ref{fig:FRA:LargeN_Fracture_Soft_Pip_BRIT}, \ref{fig:FRA:LargeN_Fracture_Soft_Pip_DUCT},  \ref{fig:FRA:LargeN_Fracture_Soft_Pip_OVERDUCT}, where the same quantities were represented for large, but finite values of $N$. 

\begin{figure*}
    \centering
    \includegraphics[scale=1]{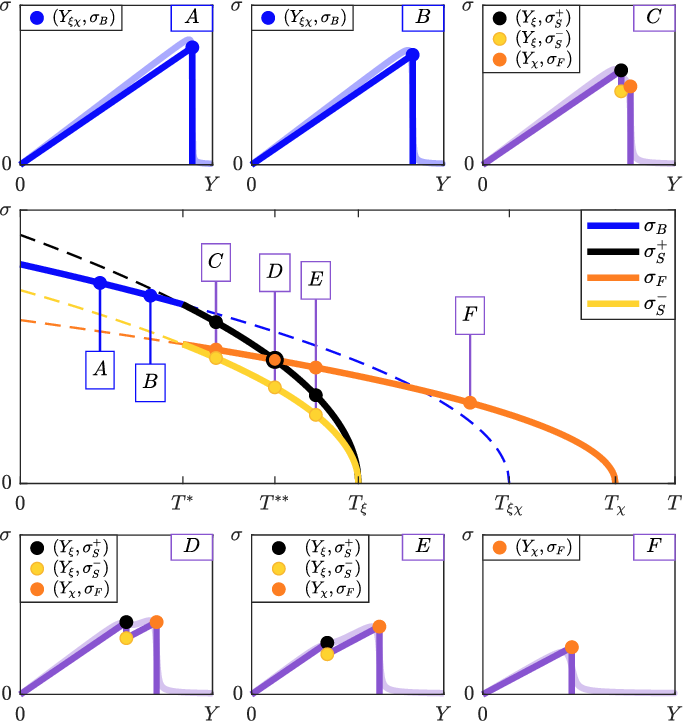}
    \caption{Behavior of the characteristic stresses $\sigma_B$, $\sigma_S^+$, $\sigma_S^-$, and $\sigma_F$ versus the temperature $T$, as defined in Eqs.(\ref{stressB}), (\ref{stressS+}), (\ref{stressS-}) and (\ref{stressF}) with $\phi=0$. Moreover, some stress-extension curves are plotted in correspondence of the following values of the temperature: A) $0<T<T^*$; B) $T=(T^*)^-$ (on the left of $T^*$); C) $T=(T^*)^+$ (on the right of $T^*$); D) $T=T^{**}$; E) $T^{**}<T<T_\xi$; F) $T_\xi<T<T_\chi$. While the temperature $T^*$ indicates the switching between brittle and ductile behavior, see Eq.(\ref{tempaste}), the temperature $T^{**}$ corresponds to $\sigma_S^+=\sigma_F$, see Eq.(\ref{tempdueaste}). In panels A), B), C), D), E), and F) we also show the stress-extension response for $N=1000$ (dim colors or light gray curves).}
    \label{fig:fra:infNstress_soft}
\end{figure*}

These results allow us to identify the values of stress corresponding to the behavioral transitions.
In the brittle regime ($0<T<T^*$) we identify the fracture or breaking stress corresponding to $\sigma_B=\lim_{N\to\infty}{\langle f \rangle}/{N}$, for $Y=Y_{\xi\chi}$, which assumes the value
\begin{eqnarray}
\nonumber
 \sigma_B&=&(1-\phi)\sqrt{k\beta^2\left(\frac{1}{\beta}-\frac{1}{\alpha} \right)\left[pY_b^2+(h-p)Y_p^2\right]\left(1-\frac{T}{T_{\xi\chi}}\right)}\\
  \label{stressB}
 &=&(1-\phi)\sqrt{\frac{lh}{l+h}\left[pY_b^2+(h-p)Y_p^2\right]\left(1-\frac{T}{T_{\xi\chi}}\right)},
\end{eqnarray}
depending on the critical temperature $T_{\xi\chi}$ (see fourth row, first panel, of Fig.\ref{fig:fra:infN_soft}).
In the ductile regime ($T^*<T<T_\xi$), we have a first transition coinciding with the softening of all elements. It represents the beginning  of the plastic regime. Two different values of stress describe this transition: the upper yield strength $\sigma_S^+=\lim_{N\to\infty}{\langle f \rangle}/{N}$ (for $Y=Y_{\xi}^-$, i.e. on the left of $Y_{\xi}$), and the lower yield strength $\sigma_S^-=\lim_{N\to\infty}{\langle f \rangle}/{N}$ (for $Y=Y_{\xi}^+$, i.e. on the right of $Y_{\xi}$), given by
\begin{eqnarray}
    \nonumber
    \sigma_S^+&=&(1-\phi)\sqrt{k\beta^2\frac{\left(\frac{1}{\beta}-\frac{1}{\alpha} \right)^2}{\frac{1}{\gamma}-\frac{1}{\alpha} }(h-p)Y_p^2\left(1-\frac{T}{T_{\xi}}\right)}\\
    \label{stressS+}
    &=&(1-\phi)\sqrt{h^2\frac{l+p}{l+h}Y_p^2\left(1-\frac{T}{T_{\xi}}\right)},
\end{eqnarray}
\begin{eqnarray}
    \nonumber
    \sigma_S^-&=&(1-\phi)\sqrt{k\beta^2\frac{\left(\frac{1}{\beta}-\frac{1}{\gamma} \right)^2}{\frac{1}{\gamma}-\frac{1}{\alpha} }(h-p)Y_p^2\left(1-\frac{T}{T_{\xi}}\right)}\\
    \label{stressS-}
    &=&(1-\phi)\sqrt{p^2\frac{l+h}{l+p}Y_p^2\left(1-\frac{T}{T_{\xi}}\right)},
\end{eqnarray}
which depend on the critical temperature $T_{\xi}$ (see fourth row, second panel, of Fig.\ref{fig:fra:infN_soft}).
These two values are useful to calculate the stress jump corresponding to the softening mechanism (yielding) within the ductile regime
\begin{eqnarray}
    \sigma_S^+-\sigma_S^-&=&(1-\phi)Y_p\frac{l(h-p)}{\sqrt{(l+h)(l+p)}}\sqrt{1-\frac{T}{T_{\xi}}},\,\,\,\,\,\,
\end{eqnarray}
which is always positive since $h>p$.
Still in the ductile regime ($T^*<T<T_\xi$), we observe the second transition describing  the complete failure of all the elements for a stress $\sigma_F=\lim_{N\to\infty}{\langle f \rangle}/{N}$, for $Y=Y_{\chi}$, assuming the value
\begin{eqnarray}
    \nonumber
    \sigma_F&=&(1-\phi)\sqrt{k\left(\frac{1}{\beta}-\frac{1}{\gamma} \right)p\left(1-\frac{T}{T_{\chi}}\right)}\beta\, Y_b\\
    \label{stressF}
    &=&(1-\phi)\sqrt{\frac{l}{l+p}\left(1-\frac{T}{T_{\chi}}\right)}p\, Y_b^2,
\end{eqnarray}
depending on the critical temperature $T_{\chi}$ (see fourth row, second panel, of Fig.\ref{fig:fra:infN_soft}).
In the over-ductile regime ($T_\xi<T<T_\chi$), the complete breaking of the system occurs at the same stress $\sigma_F$ given in Eq.(\ref{stressF}) and shown in the fourth row, third panel, of Fig.\ref{fig:fra:infN_soft}.

The behavior of these transition stresses is summarized in Fig.\ref{fig:fra:infNstress_soft}, where they are plotted versus the temperature $T$. 
In addition, different stress-extension curves are shown at different temperatures of interest. 
In the first two cases, A) and B), we observe a brittle behavior characterized by the breaking of the system when the stress reaches the value $\sigma_B$ and the extension the value $Y_{\xi\chi}$. 
While the case B) corresponds to a temperature slightly smaller than $T^*$ (brittle), the case C) represents a temperature slightly larger than $T^*$, being therefore in the ductile region. 
We see here both the softening transition at $Y_\xi$ and the failure transition at $Y_\chi$. 
In this case C), the stresses satisfy the relationship $\sigma_S^-<\sigma_F<\sigma_S^+$ and then the softening peak (upper yield strength) is larger than the failure peak. 
We can now continue to increase the temperature until  $\sigma_S^-<\sigma_F=\sigma_S^+$, that is, until the softening peak is equal to the failure peak.
This condition is fulfilled in the panel D) of Fig.\ref{fig:fra:infNstress_soft}, and it corresponds to the temperature $T^{**}$, defined as
\begin{equation}
    T^{**} = \frac
    {\frac{h^2}{l+h}Y_p^2-\frac{p^2l}{(p+l)^2}Y_b^2}
    {K_B\left[\frac{h^2}{(l+h)(h-p)}\ln\frac{\tau_\alpha}{\tau_\gamma}-\frac{pl}{(p+l)^2}\ln\frac{\tau_\gamma}{\tau_\beta}\right]}.
    \label{tempdueaste}
\end{equation}
In the stress-temperature plot, this temperature value $T^{**}$ represents the intersection of the two curves $\sigma_F$ and $\sigma_S^+$ versus $T$.
Increasing the temperature further, we enter the region $T^{**}<T<T_\xi$ (once again ductile), represented in the panel E), where the failure peak is larger than the softening peak (upper yield strength), $\sigma_S^-<\sigma_S^+<\sigma_F$. 
Finally, for values of temperature in the range $T_\xi<T<T_\chi$, we are in the over-ductile regime and the softening peak disappears, remaining only the failure peak $\sigma_F$ for the overall system, as shown in panel F) of Fig.\ref{fig:fra:infNstress_soft}. 
In panels A), B), C), D), E), and F) of Fig.\ref{fig:fra:infNstress_soft} we also represent the stress-extension response for a finite (large) value of $N$ in order to show the good agreement between the approximated expressions and the thermodynamic limit.

As a conclusion to this discussion, we would like to point out that the strength (rupture stress) of the system as a function of temperature is finally represented by a discontinuous curve formed by the branch $\sigma_B$ for $0<T<T^*$ (brittle) and by the branch $\sigma_F$ for $T^*<T<T_\chi$ (ductile), as one can see in the panel $\sigma-T$ of  Fig.\ref{fig:fra:infNstress_soft}. 
This discontinuity can be easily explained by observing that the brittle-to-ductile transition involves the phenomenon of softening and thus the synchronized lowering of the elastic constant of all breakable elements. Since we are applying a stretching to the system controlled by the extension, the reduction of the overall elastic constant produces a consequent reduction in stress (which is therefore discontinuous). 
We further remark that the strength ($\sigma_B$ or $\sigma_F$, depending on the temperature) is proportional to the factor $1-\phi$, which represents the fraction of initially present elements ($\phi$ is in fact the fraction of initially absent elements). This is reminiscent of the Griffith criterion, stating that the stress at fracture is lower if the initial crack opening is larger \cite{griffith}. In our case, the initial crack opening is proportional to $\phi$ and, therefore, the Griffith criterion is respected. However, we add here the temperature dependent nature of this criterion, which is described by the classical term $\sqrt{1-\frac{T}{T_{\xi\chi}}}$ in $\sigma_B$, or $\sqrt{1-\frac{T}{T_{\chi}}}$ in $\sigma_F$, which represents the critical behavior eventually resulting in a genuine phase transition.

\begin{figure}
    \centering
    \includegraphics[scale=1]{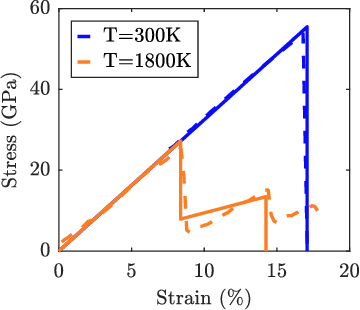}
    \caption{Tensile stress-strain curves for a GaN nanowire oriented in the direction  [0001] (diameter of 1.92 nm, length of 6.12 nm). The lateral facets are oriented along the $\left\lbrace 11\bar{2}0\right\rbrace $ side planes. Comparison between molecular dynamics simulations results \cite{weber}  (dashed lines) and our theory given by Eqs.(\ref{brittlestress}) and (\ref{ductilestress}) (continuous lines). The parameters used are reported in the main text.}
    \label{fit-brittle-ductile}
\end{figure}

The brittle-to-ductile transition has been observed in GaN nanowires through molecular dynamics simulations \cite{weber}, and Fig.\ref{fit-brittle-ductile} shows the comparison with our theoretical results. We considered a GaN nanowire oriented in the direction  [0001], with a diameter of 1.92 nm and a length of 6.12 nm, as reported in Fig.2(d) of Ref.\cite{weber}. The lateral facets of this system are oriented along the $\left\lbrace 11\bar{2}0\right\rbrace $ side planes, as shown in Fig.1(b) of Ref.\cite{weber}. In Fig.\ref{fit-brittle-ductile}, the blue curves (or dark gray) represent the brittle behavior whereas the orange (or light gray) ones describe the ductile behavior. 
We remark that Eqs.(\ref{brittlestress}) and (\ref{ductilestress}) define a relation $\left\langle f\right\rangle /N=\mathcal{F}(Y)$ where $\mathcal{F}$ is a given function. We have to introduce the real stress $\sigma=\left\langle f\right\rangle /(N\mathsf{S})$, where  $\mathsf{S}$ is the area pertaining to each breakable spring, and the real strain $\varepsilon=Y/\ell$, where $\ell$ is the characteristic lengthscale induced by the crystal structure. 
The stress-strain relation can be therefore written as $\sigma=\mathcal{F}(\varepsilon\ell)/\mathsf{S}$, where $\mathcal{F}$ is the relationship defined by Eqs.(\ref{brittlestress}) and (\ref{ductilestress}).
In Fig.\ref{fit-brittle-ductile}, we adopted the parameters $h    = 3.77$N/m, $p    = 0.234$N/m, $l = 0.725$N/m, $k = 2.00$N/m, $\Delta E   = 11.0\times 10^{-21}$J, $Y_b  = 17.8\times 10^{-11}$m, $\mathsf{S}    = 22.7\times 10^{-22}$m$^2$, $\ell = 12.1\times 10^{-12}$m, and $K_B = 1.38\times 10^{-23}$J/K. 
As before, most of geometrical parameters were available in the original paper dealing with molecular dynamics simulations and the others were fitted to correctly reproduce the results.
We then plotted the stress-strain curves for the two temperatures $T=300$K (brittle behavior) and $T=1800$K (ductile behavior).
It is interesting to note that the structural parameters used in our model are able to predict the correct brittle-to-ductile transition as obtained through  molecular dynamics simulations. Moreover, also the upper and lower yield stresses of the ductile behavior are in quite good agreement with simulations. We remark that in our model there is a single softening process and therefore we can see only one failure peak after the softening peak in the stress-strain relation. As discussed in the conclusions, the model could be generalized with more softening steps to describe real damage such as that of the nanowires studied here.

\section{Conclusions}

We proposed and studied two prototypical models  able to describe temperature effects in fracture processes. 
The first is aimed at explaining the temperature dependent behavior for brittle systems, and the second at showing the complex damage processes occurring in the presence of ductile breakable links with a possibility of a brittle-to-ductile transition regulated by thermal effects.   
Both models are based on a simple lattice structure built through unbreakable and breakable springs.  
The system is confined between two layers, one fixed and one movable, and is designed in such a way that lifting the top layer results in a force experienced by the system, being able to generate fracture propagation. 
This structure is supposed to be embedded into a thermal bath at fixed temperature. 
Hence, the models are developed within the equilibrium statistical mechanics formalism.  
The difference between the two models lies in the behavior at rupture of the breakable springs. 
In the first model for brittle fracture, each breakable spring can be in two states, namely elastic or broken, depending on the extension applied to the element itself. 
The state transition occurs through the absorption of an amount of energy that corresponds to the typical surface energy of the Griffith criterion \cite{griffith}. 
In the second model, each breakable spring can be in three different states representing the elastic, softened, and broken regimes. 
The intermediate softened state is introduced to reproduce the possible ductile regime of the fracture process. 
In this case, we have a `yielding' point between the elastic and the ductile regimes, followed by a final failure point corresponding to fracture. 
The transition between intact and softened states occurs after the yielding energy is absorbed, and the transition between softened and broken occurs through the absorption of another amount of energy corresponding to fracture.
Thus, in both models the energy balance is similar to what is typically assumed in linear elastic fracture mechanics since Griffith's and Irwin's  pioneering works \cite{griffith,Irwin1,Irwin2}, with the fracture phenomenon regulated by elastic, damage and fracture (surface) energy.  
However, including thermal fluctuations sensibly modifies the results and add important features to the system description. 

As for the model for brittle fracture, we obtain a temperature-dependent fracture stress and a corresponding fracture strain, representing a `genuine' phase transition. 
Thus, we obtain a critical temperature at which both fracture stress and fracture strain are zero and therefore the material is always broken for supercritical temperatures.
Interestingly, the obtained temperature-dependent strength is in good agreement with several experiments and molecular dynamics simulations as demonstrated previously.
It is interesting to note that although breakable springs have a temperature-independent breaking behavior, the overall system exhibits a breaking point that is highly dependent on temperature. This is a typical case of a complex system with collective behavior, giving rise to a critical phenomenon. 
We argue that this effect is relevant in the case of weak links, such as hydrogen bonds in biological materials, or in rubber, where the elasticity has an entropic character, or in small size metallic or semiconductor systems, such as the considered nanowires \cite{pennings,cai1,weber,cai2,Srolovitz,mao,maeder,saha,zhu,rubber,coarse,dna}.

The model with ductile breakable elements exhibits an even richer behavior. 
In this case it is the full response of the system that is temperature dependent. 
Indeed, we have demonstrated the existence of a brittle-to-ductile  transition temperature $T^*$ (whose expression is obtained in closed form) that regulates the behavior of the fracture process. 
On the one hand, for temperatures lower than $T^*$, we observe a brittle behavior characterized by a direct transition of the springs from the elastic to the broken state, without passing through the intermediate softened state. On the other hand, for temperatures higher than $T^*$, we see that, as the extension applied to the system increases, first the springs soften (yielding point), and then they switch from the softened to the broken state (failure point). 
The intermediate softened region reproduces in this discrete context the cohesive zone of the classical Dugdale-Barenblatt model of the ductile fracture \cite{dugdale,barenblatt}.
Of course, both yielding and failure point depend on temperature and are again characterized by phase transitions. 
In fact, importantly, we are not only able to calculate the brittle-to-ductile transition temperature, but also to predict the critical behavior of the upper and lower yield strengths, and the thermal properties of the fracture strength. 
Our model is also able to predict the existence of a special fracture regime, here called over-ductile, in which the temperature is high enough to damage all elements without mechanical action.  In this situation, as the extension of the system increases, we observe the only transition from the softened to the broken regime. 

From the methodological point of view, to elaborate the closed form expression of the partition function in both proposed models, we adopted specific techniques particularly suitable for calculating the determinant and inverse of tridiagonal matrices \cite{usmani1,usmani2}. 
These approaches allow the derivation of exact solutions as shown in Appendix \ref{appendixa}, but also asymptotic approximations as discussed in Appendix \ref{appendixb}. Although these mathematical developments are relegated to the appendices, they are of crucial importance for obtaining the physical results on fracture processes. 

 We point out that even if the models here presented  clarify fundamental aspects of thermally activated rupture phenomena, they should be generalized to take into account the complex reality of these processes. We want to mention here at least four points that partially limit the applicability of these models to real situations.
The first issue concerns the spatial homogeneity of the adopted models. We have always considered all springs of discrete systems having the same mechanical behavior (in terms of elastic constants, failure thresholds, etc.). 
In reality, this is true only for perfect monocrystalline structures that are quite rare. It would be interesting to study these phenomena in disordered systems that, on the one hand, are more similar to several real structures, and on the other hand, may generate even more interesting critical behaviors typical of complex systems with quenched disorder \cite{diso,parisi,ponson,niccolini,sinha,kondo,rocha}. 
The second point to be explored is the kinetic of rupture processes. Here we have considered only quasi-static phenomena studied by means of equilibrium statistical mechanics. In real experiments, traction can be applied at different tensile velocities, and the response obviously depends on these traction rates \cite{chen,bonamy}.
To model these phenomena one would have to adopt out-of-equilibrium statistical mechanics  and then base the analysis on Langevin or Fokker-Planck methodologies \cite{risken,coffey,ben1,ben2}.
To conclude, the third point that could be improved concerns the fact that the softening process is restricted to a single step of reduction of the elastic constant of the breakable springs. 
In order to be more adherent to the physical reality of the yielding process one would have to imagine a series of steps where several reductions of the elastic constant take place progressively. 
In this sense, the yielding point would be implemented through a multi-softening process, more similar to what happens in real nonlinear materials.
The fourth and final point concerns the too-simple geometry of our model, which should be improved (with 2D or 3D lattices) in order to be able to represent real elastic fields to be compared with models from continuum mechanics. 

\appendix

\section{Exact results for tridiagonal matrices}
\label{appendixa}

Since the matrix $\mathcal{A}$ defined in Eq.(\ref{eq:fra:tridiagonal_matrix}) is  tridiagonal, we can analytically evaluate the inverse $\mathcal{A}^{-1}$ and the determinant $\det \mathcal{A}$ \cite{usmani1,usmani2}.
We consider  a generic tridiagonal matrix $\mathcal{M}$, and we define its elements  as $\mathcal{M}_{i,i} = b_i$ (main diagonal), $\mathcal{M}_{i,i-1} = a_i$ (lower diagonal), and $\mathcal{M}_{i,i+1} = c_i$ (upper diagonal). All other elements are zero. We can introduce the quantities $\theta_i$ by means of the following recurrence relation
\begin{equation}
\label{eq:fra:thetas}
    \begin{cases}
        \theta_i = b_i\theta_{i-1}-a_ic_{i-1}\theta_{i-2},\\
        \theta_{-1}=0,\,\theta_{0}=1,\,i=1,2,\dots,N,
    \end{cases}
\end{equation}
where, in particular, $\theta_N = \det\mathcal{M}$.
Furthermore, it is possible to define the quantities $\phi_i$ through the recurrence formula
\begin{equation}
\label{eq:fra:phis}
    \begin{cases}
        \phi_i = b_i\phi_{i+1}-c_ia_{i+1}\phi_{i+2},\\
        \phi_{N+2}=0,\,\phi_{N+1}=1,\,i=N,N-1,\dots,1,
    \end{cases}
\end{equation}
where $\phi_1=\theta_N=\det\mathcal{M}$.
These definitions can be used to determine the elements of the inverse matrix $\mathcal{M}^{-1}$ \cite{usmani1,usmani2}, as follows
\begin{equation}
    (\mathcal{M}^{-1})_{i,j} = \left\{
    \begin{aligned}
        &\frac{(-1)^{i+j}c_ic_{i+1}\dots c_{j-1}\theta_{i-1}\phi_{j+1}}{\theta_N},&\quad\text{if }i<j,\\
        &\frac{\theta_{i-1}\phi_{i+1}}{\theta_N},&\quad\text{if }i=j,\\
        &\frac{(-1)^{i+j}a_{j+1}a_{j+2}\dots a_i\theta_{j-1}\phi_{i+1}}{\theta_N},&\quad\text{if }i>j.
    \end{aligned}
    \right.
\end{equation}
By considering our particular case, the elements in the main diagonal of $\mathcal{A}$ are defined as $a_i = 2+\alpha$ for $1\leq i\leq\xi$, and $a_i = 2+\beta$ for $\xi +1\leq i\leq N$. Moreover, we have that $\mathcal{A}_{i,i+1} = \mathcal{A}_{i+1,i} = -1$ for the upper and lower diagonals. For this special situation, $\theta_i$ and $\phi_i$ are defined by the rules
\begin{equation}
    \begin{cases}
    \label{eq:fra:thetas_A}
        \theta_i = a_i\theta_{i-1}-\theta_{i-2},\\
        \theta_{-1}=0,\,\theta_{0}=1,\,i=1,2,\dots,N,
    \end{cases}
    \end{equation}
and
\begin{equation}
    \begin{cases}
    \label{eq:fra:phis_A}
        \phi_i = a_i\phi_{i+1}-\phi_{i+2},\\
        \phi_{N+2}=0,\,\phi_{N+1}=1,\,i=N,N-1,\dots,1.
    \end{cases}
\end{equation}
Consequently, the elements of the inverse matrix $\mathcal{A}^{-1}$ are given by
\begin{equation}
    \label{eq:fra:inverse_components}
    (\mathcal{A}^{-1})_{i,j} = \left\{
    \begin{aligned}
        \frac{\theta_{i-1}\phi_{j+1}}{\theta_N},&\quad\text{if }i<j,\\
        \frac{\theta_{i-1}\phi_{i+1}}{\theta_N},&\quad\text{if }i=j,\\
        \frac{\theta_{j-1}\phi_{i+1}}{\theta_N},&\quad\text{if }i>j.\\
    \end{aligned}
    \right.
\end{equation}
Hence, we need to find $\theta_i$ and $\phi_i$ in order to obtain the inverse matrix elements.
We start by evaluating $\theta_i$ for $i\leq\xi$.
In this case, $a_i = 2 + \alpha$, and Eq.\eqref{eq:fra:thetas_A} becomes
\begin{equation}
\label{eq:fra:theta_alpha}
    \theta_i = (2+\alpha)\theta_{i-1}-\theta_{i-2}.
\end{equation}
To find a solution, we substitute $\theta_i = \lambda^i$ in the last equation, and we obtain a second degree algebraic equation with solutions
\begin{equation}
    \lambda_{1,2} = \frac{2+\alpha\pm\sqrt{\alpha^2-4\alpha}}{2}.
\end{equation}
Then, a generic solution for $\theta_i$, with $i\leq\xi$, is given by the following linear combination 
\begin{equation}
    \theta_i = A\left(\frac{2+\alpha+\sqrt{\Delta_\alpha}}{2}\right)^i+B\left(\frac{2+\alpha-\sqrt{\Delta_\alpha}}{2}\right)^i,
\end{equation}
where we introduced $\Delta_\alpha = \alpha^2+4\alpha$.
We can obtain the two coefficients $A$ and $B$ by the initial conditions in Eq.\eqref{eq:fra:thetas_A}. 
We  obtain 
\begin{equation}
    A = \frac{2+\alpha+\sqrt{\Delta_\alpha}}{2\sqrt{\Delta_\alpha}},\quad B = -\frac{2+\alpha-\sqrt{\Delta_\alpha}}{2\sqrt{\Delta_\alpha}}.
\end{equation}
Therefore, the final solution for $\theta_i$, when $i\leq\xi$, is
\begin{equation}
    \theta_i = \mathcal{G}(\alpha,i+1),
\end{equation}
where we introduced the function
\begin{equation}
    \label{eq:fra:G_definition}
    \mathcal{G}(\gamma, z) = \frac{1}{\sqrt{\Delta_\gamma}}\left[\left(\frac{2+\gamma+\sqrt{\Delta_\gamma}}{2}\right)^z-\left(\frac{2+\gamma-\sqrt{\Delta_\gamma}}{2}\right)^z\right],
\end{equation}
with $\Delta_\gamma = \gamma^2+4\gamma$. 
If we introduce the parameters $\tau_\gamma$ and $\rho_\gamma$ as follows
\begin{equation}
            \tau_\gamma = \frac{2+\gamma+\sqrt{\Delta_\gamma}}{2},\quad
        \rho_\gamma = \frac{2+\gamma-\sqrt{\Delta_\gamma}}{2},
  \end{equation}
the function $\mathcal{G}(\gamma, z)$ can be written as
\begin{equation}
    \mathcal{G}(\gamma, z) = \frac{1}{\sqrt{\Delta_\gamma}}\left(\tau_\gamma^z-\rho_\gamma^z\right).
\end{equation}
We note that 
\begin{equation}
    \label{eq:fra:tau_rho_properties}
        \tau_\gamma-\rho_\gamma=\sqrt{\Delta_\gamma},\quad
        \rho_\gamma\tau_\gamma=1.
\end{equation}
Now, we evaluate $\theta_i$ when $i\geq \xi+1$.
In this case, $a_i = 2 + \beta$ and Eq.\eqref{eq:fra:thetas_A} becomes
\begin{equation}
    \theta_i = (2+\beta)\theta_{i-1}-\theta_{i-2}.
\end{equation}
As before, we find that the general solution is given by the linear combination
\begin{equation}
    \theta_i = C\left(\frac{2+\beta+\sqrt{\Delta_\beta}}{2}\right)^i+D\left(\frac{2+\beta-\sqrt{\Delta_\beta}}{2}\right)^i,
\end{equation}
where $\Delta_\beta = \beta^2+4\beta$.
To find the coefficients $C$ and $D$, we exploit the initial conditions $\theta_\xi = \mathcal{G}(\alpha,\xi+1)$ and $\theta_{\xi-1} = \mathcal{G}(\alpha,\xi)$. Straightforward calculations lead to the solution for $\theta_i$, when $i\geq \xi+1$, in the form
\begin{equation}
    \theta_i = \mathcal{G}(\beta,i-\xi+1)\mathcal{G}(\alpha,\xi+1)-\mathcal{G}(\beta,i-\xi)\mathcal{G}(\alpha,\xi),
\end{equation}
where we used the function defined in Eq.(\ref{eq:fra:G_definition}).
We consider Eq.\eqref{eq:fra:phis_A} and we proceed with the evaluation of $\phi_i$.
We start with the case where $i\geq \xi+1$. In this condition, the recurrent equation becomes
\begin{equation}
    \phi_i = (2+\beta)\phi_{i+1}-\phi_{i+2},
\end{equation}
which must be combined with the initial conditions in Eq.\eqref{eq:fra:phis_A}. Eventually, we obtain $\phi_i$ for $i\geq \xi+1$ in the form
\begin{equation}
    \phi_i = -\mathcal{G}(\beta,i-N-2).
\end{equation}
We can find $\phi_i$ when $i\leq\xi$ by using the two conditions $\phi_{\xi+1}=-\mathcal{G}(\beta,\xi-N-1)=\mathcal{G}(\beta,N+1-\xi)$ and $\phi_{\xi+2}=-\mathcal{G}(\beta,\xi-N)=\mathcal{G}(\beta,N-\xi)$.
After straightforward calculations, we get for $i\leq\xi$
\begin{eqnarray}
\nonumber
    \phi_i &=& \mathcal{G}(\beta,N+1-\xi)\mathcal{G}(\alpha,\xi+2-i)\\
    &&-\mathcal{G}(\beta,N-\xi)\mathcal{G}(\alpha,\xi+1-i).
\end{eqnarray}
The obtained values of $\theta_i$ and $\phi_i$ allow the calculation of $\mathcal{A}^{-1}$ and $\det\mathcal{A}$, useful to implement the determination of the partition function in Eq.(\ref{eq:fra:HpartfunQ}) and the quantities  in Eqs.(\ref{eq:fra:f-e_Helm2}) and (\ref{eq:fra:xi_Helm2}).
Moreover, these results are useful to develop some asymptotic expressions in Appendix \ref{appendixb}.

\section{Asymptotic analysis}
\label{appendixb}

Considering the brittle model with a large number $N$ of units, it is possible to derive approximations to simplify the partition function and the main average quantities. More specifically, we can find approximations for $\vec{v}\cdot\mathcal{A}^{-1}\vec{v}$ and for $\det\mathcal{A}$.
We start our analysis by expanding the first quadratic form as follows
\begin{equation}\label{eq:fra:vtav}
    \begin{aligned}
        \vec{v}\cdot\mathcal{A}^{-1}\vec{v} =& \sum_{i=1}^N\sum_{j=1}^Nv_i(\mathcal{A}^{-1})_{i,j}v_j\\
        =&\sum_{i=1}^N\sum_{j=1}^N(\beta+\delta_{i,N})(\mathcal{A}^{-1})_{i,j}(\beta+\delta_{j,N})\\
        =&\beta^2\sum_{i=1}^N\sum_{j=1}^N(\mathcal{A}^{-1})_{i,j}+2\beta\sum_{i=1}^N(\mathcal{A}^{-1})_{i,N}+(\mathcal{A}^{-1})_{N,N}\\
        =&\beta^2S_2(\xi)+2\beta S_1(\xi)+S_0(\xi),
    \end{aligned}
\end{equation}
where we introduced $S_2(\xi)$ as the sum over all the elements of the inverse matrix, $S_1(\xi)$ as the sum over all the elements of the $N$-th column of the inverse matrix, and $S_0(\xi)$ as the  element $(N, N)$ of the inverse matrix.
We observe that these three quantities are in general function of $\xi$.

Exploiting the symmetry of the inverse matrix $\mathcal{A}^{-1}$, we write $S_2(\xi)$ as
\begin{equation}
    \label{eq:fra:addend}
    S_2(\xi) = \sum_{i=1}^N(\mathcal{A}^{-1})_{i,i} + 2 \sum_{i=1}^{N-1}\sum_{j=i+1}^N(\mathcal{A}^{-1})_{i,j}.
\end{equation}
The evaluation of $S_2(\xi)$ for a matrix $\mathcal{A}$ that shows heterogeneous diagonal elements $a_i$ ($\xi\neq\{0,N\}$) can be done but is not straightforward.
Fortunately, it is possible to observe that, in the limit of large $N$, the form of $S_2(\xi)$ is a linear combination of the two values $S_2(0)$ and $S_2(N)$, each corresponding to a matrix with homogeneous diagonal.
In fact, when $\xi=0$, the diagonal components of $\mathcal{A}$ are all equal to $2+\beta$, and when $\xi=N$, the diagonal components of $\mathcal{A}$ are all equal to $2+\alpha$.

\begin{figure}
    \centering
    \includegraphics[scale=1]{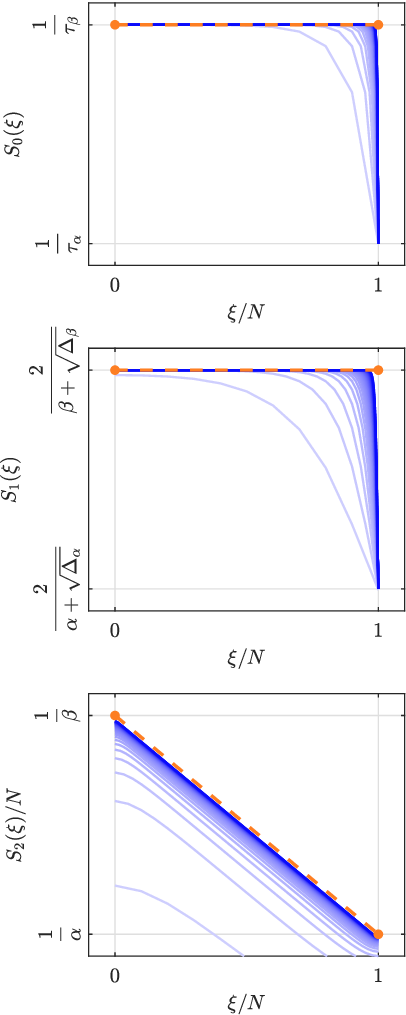}
    \caption{The quantities $S_2(\xi)/N$, $S_1(\xi)$ and $S_0(\xi)$ are obtained using Usmani relations (continuous curves) for different values of $N=\{10,\dots,300\}$ (with step of 10). We observe that, as $N$ increases, the quantities approach their relative approximations obtained for large $N$ (dashed lines).}
    \label{fig:FRA:Tridiagonal_vAv}
\end{figure}

With the help of the left panel of Fig.\ref{fig:FRA:Tridiagonal_vAv}, we can observe that, as $N$ increases, the form of $S_2(\xi)$ numerically obtained through Eq.\eqref{eq:fra:inverse_components} (continuous curves), approach the straight line joining $S_2(0)$ and $S_2(N)$ (dashed line).
To improve the approximation, we observe that in the linear solution for $S_2(\xi)$ we could also add a zeroth order term (with respect to $N$), represented by an additional small quantity $C(\xi)$, possibly dependent on $\xi$, but independent of $N$.

Now, we analytically evaluate $S_2(0)$. By means of this value, it is also easy to obtain $S_2(N)$ simply substituting $\beta$ with $\alpha$.
When $\xi=0$, we have from Appendix \ref{appendixa} 
\begin{align}
    &\theta_i = \mathcal{G}(\beta, i+1),\\
    &\phi_i = \mathcal{G}(\beta, N+2-i),
\end{align}
where $i=1,\dots,N$. In addition, Eq.\eqref{eq:fra:inverse_components} gives
\begin{equation}
    (\mathcal{A}^{-1})_{i,j} = \left\{
    \begin{aligned}
        \frac{\mathcal{G}(\beta, i)\mathcal{G}(\beta, N+1-j)}{\mathcal{G}(\beta, N+1)},\quad\text{if }i<j,\\
        \frac{\mathcal{G}(\beta, i)\mathcal{G}(\beta, N+1-i)}{\mathcal{G}(\beta, N+1)},\quad\text{if }i=j,\\
        \frac{\mathcal{G}(\beta, j)\mathcal{G}(\beta, N+1-i)}{\mathcal{G}(\beta, N+1)},\quad\text{if }i>j.\\
    \end{aligned}
    \right.
\end{equation}
We can therefore write $S_2(0)$ as
\begin{eqnarray}
\nonumber
    S_2(0) &=& \sum_{i=1}^N\frac{\mathcal{G}(\beta, i)\mathcal{G}(\beta, N+1-i)}{\mathcal{G}(\beta, N+1)} \\&&+ 2\sum_{i=1}^{N-1}\sum_{j=i+1}^N\frac{\mathcal{G}(\beta, i)\mathcal{G}(\beta, N+1-j)}{\mathcal{G}(\beta, N+1)}.
\end{eqnarray}
Using the definition of $\mathcal{G}(\gamma, z)$  in Eq.\eqref{eq:fra:G_definition}, and the properties of $\tau_\gamma$ and $\rho_\gamma$ introduced in Eq.\eqref{eq:fra:tau_rho_properties},  we get
\begin{equation}
    \label{eq:fra:long_S2}
    \begin{aligned}
        S_2(0) =& \frac{(\tau_\beta^{N+1}-\tau_\beta^{-N-1})^{-1}}{\sqrt{\Delta_\beta}}\\
        &\times\Bigg\{
        2(N-1)\left(\frac{\tau_\beta^{N+1}}{\tau_\beta-1}-\frac{\tau_\beta^{-N}}{\tau_\beta-1}\right)+\frac{2\tau_\beta^{2-N}-2\tau_\beta^N}{\tau_\beta^2-1}\\
        &+N(\tau_\beta^{N+1}+\tau_\beta^{-N-1})+\frac{2\tau_\beta}{(\tau_\beta^2-1)}(\tau_\beta^{-N}-\tau_\beta^N)\\
        &+\frac{2+2\tau_\beta}{(\tau_\beta-1)^2}(1+\tau_\beta-\tau_\beta^N-\tau_\beta^{1-N})
    \Bigg\},
    \end{aligned}
\end{equation}
where we used several times the geometric sum.
The expression for $S_2(0)$ given in Eq.\eqref{eq:fra:long_S2} is not transparent but, in the limit of large $N$, it can be approximated by 
\begin{equation}
    S_2(0) \sim \frac{N}{\beta} - \frac{\sqrt{\beta^2+4\beta}-\beta}{\beta^2} = \frac{N}{\beta} + C(0),
\end{equation}
where
\begin{equation}
\label{eq:fra:c0}
    C(0) = - \frac{\sqrt{\beta^2+4\beta}-\beta}{\beta^2}.
\end{equation}
This term represents the zeroth order correction (with respect to $N$), previously discussed.
When $\xi=N$, we can obtain the result by simply  substituting $\beta$ with $\alpha$, eventually obtaining
\begin{equation}
    S_2(N) \sim \frac{N}{\alpha} - \frac{\sqrt{\alpha^2+4\alpha}-\alpha}{\alpha^2} = \frac{N}{\alpha} + C(N),
\end{equation}
where
\begin{equation}
\label{eq:fra:cN}
    C(N) = - \frac{\sqrt{\alpha^2+4\alpha}-\alpha}{\alpha^2}.
\end{equation}
Finally, we can write the general approximation for $S_2(\xi)$ in the limit of large $N$ as
\begin{equation}
    S_2(\xi) \sim \frac{N}{\beta}+\left(\frac{N}{\alpha}-\frac{N}{\beta}\right)\frac{\xi}{N}+C(\xi),\quad\text{if }\alpha,\beta\neq 0,
\end{equation}
where $C(0)$ is given in Eq.\eqref{eq:fra:c0}, $C(N)$ is given in Eq.\eqref{eq:fra:cN}, and $C(\xi)$ assumes a constant value for $\xi \in \{1,\dots,N-1\}$ (for large $N$), which is always in the range between $C(0)$ and $C(N)$.
We do not determine here this value since is not relevant for our analysis.
Indeed, although the zeroth order term of $S_2(\xi)$ is represented by three different values of the constant depending on $\xi$, in the application to the fracture problem we  adopt the value $C(0)$ in all calculations.
It is not difficult to realize that this is the only value playing a role in our model since it describes the behavior of the system when $\xi=0$, i.e. when all the breakable springs are fractured.
In this condition, only one spring links together the two layers of the system and the constant $C(0)$ is able to describe the exact stiffness of the resulting spring network.
The other values $C(\xi)$, for $\xi\neq 0$, are negligible when $N\to\infty$.

The approach used to find the approximation of $S_2(\xi)$ for large $N$, can be also applied for $S_1(\xi)$ and $S_0(\xi)$.
Concerning $S_1(\xi)$, as shown in the center panel of Fig.\ref{fig:FRA:Tridiagonal_vAv}, its value numerically obtained with Eq.\eqref{eq:fra:inverse_components} approaches the constant value $S_1(0)$ as $N$ increases.
This value is therefore the approximation of $S_1(\xi)$ for large $N$.
Through previous definitions, we can write $S_1(0)$ as
\begin{equation}
    S_1(0) = \sum_{i=1}^N(\mathcal{A}^{-1})_{i,N} = \sum_{i=1}^N\frac{\theta_{i-1}}{\theta_N} = \sum_{i=1}^N\frac{\mathcal{G}(\beta,i)}{\mathcal{G}(\beta,N+1)},
\end{equation}
which, in the limit of large $N$, leads to
\begin{equation}
    S_1(\xi) \sim S_1(0)\sim \frac{2}{\beta+\sqrt{\Delta_\beta}}.
\end{equation}
We can observe that the exact values of $S_0(\xi)$ approach the constant value $S_0(0)$ for large $N$, as one can see in the right panel of Fig.\ref{fig:FRA:Tridiagonal_vAv}. We have the exact expression
\begin{equation}
    S_0(0) = (\mathcal{A}^{-1})_{N,N} = \frac{\theta_{N-1}}{\theta_N},
\end{equation}
which, in the limit of large $N$, gives
\begin{equation}
    S_0(\xi) \sim S_0(0) \sim \frac{1}{\tau_\beta} = \frac{2}{2+\beta+\sqrt{\beta^2+4\beta}},
\end{equation}
as shown in the right panel of Fig.\ref{fig:FRA:Tridiagonal_vAv}.
Now we determine the value of the quadratic form in Eq.\eqref{eq:fra:vtav}, for large $N$, as
\begin{eqnarray}
        \vec{v}\cdot\mathcal{A}^{-1}\vec{v} &\sim & \beta^2\left[\frac{N}{\beta}+\left(\frac{1}{\alpha}-\frac{1}{\beta}\right)\xi + C(0)\right] \\
        \nonumber
        &+& 2\beta\left(\frac{2}{\beta+\sqrt{\beta^2+4\beta}}\right)+ \left(\frac{2}{2+\beta+\sqrt{\beta^2+4\beta}}\right).
    \end{eqnarray}
To conclude, we recall the definition of $q$, stated in Eq.\eqref{eq:fra:q}, and we get
\begin{equation}
        q = 1+\beta N-\vec{v}\cdot\mathcal{A}^{-1}\vec{v}\sim\frac{lh}{l+h}\frac{\xi}{k}+\epsilon,
        \label{qqapp}
\end{equation}
where we introduced
\begin{equation}
    \label{eq:fra:epsilon}
    \epsilon = \frac{\sqrt{\beta^2+4\beta}-\beta}{2}.
\end{equation}
Finally, Eqs.(\ref{qqapp}) and (\ref{eq:fra:epsilon}) prove Eqs.(\ref{qqtext}) and (\ref{eptext}) of the main text.

\begin{figure}
    \centering
    \includegraphics[scale=1]{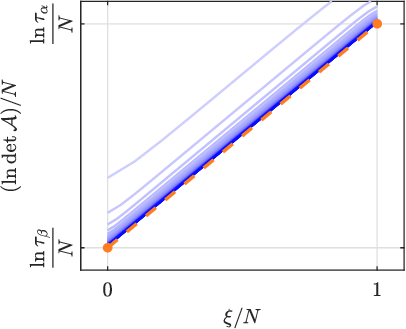}
    \caption{The quantity $(\ln\det\mathcal{A})/{N}$ is obtained using Usmani relations (continuous curves) for different values of $N=\{10,\dots,300\}$ (with steps of 10). We observe that, as $N$ increases, $(\ln\det\mathcal{A})/N$ approaches its approximation for large $N$ (dashed line).}
    \label{fig:FRA:Tridiagonal_det}
\end{figure}

To complete this part, we study the approximation of $\det\mathcal{A}$ for large values of $N$. Thanks to Usmani theory \cite{usmani1,usmani2}, we have
\begin{equation}
    \det\mathcal{A} = \theta_N.
\end{equation}
Adopting the results of Appendix \ref{appendixa}, it is possible to evaluate the determinant of $\mathcal{A}$ for different $\xi=0,\dots,N$.
The results can be found in Fig.\ref{fig:FRA:Tridiagonal_det}, from which we realize that the value of $\ln(\det\mathcal{A})/N$ is approximated by a straight line that links together the values of $\ln\det\mathcal{A}/N$ when $\xi=0$, and when $\xi=N$, in the limit of large $N$. 
It is simple to prove that for $\xi=0$ we have ${\ln\theta_N}/{N} \simeq\ln\tau_\beta$ when $N\to\infty$, and similarly for $\xi=N$ we have ${\ln\theta_N}/{N} \simeq\ln\tau_\alpha$ when $N\to\infty$.
The equation that gives the value of $\ln\det\mathcal{A}/N$, in the limit of large $N$, is therefore obtained as
\begin{equation}
    \frac{\ln\det\mathcal{A}}{N} = \frac{\ln\theta_N}{N} \sim\ln\tau_\beta + \frac{\xi}{N}\ln\frac{\tau_\alpha}{\tau_\beta}.
\end{equation}
Equivalently, we can write
\begin{equation}
            \det\mathcal{A} \sim \tau_\alpha^\xi\tau_\beta^{N-\xi},
\end{equation}
which proves Eq.(\ref{dettext}) of the main text.

We discuss now the same results for the model with the softening mechanism. In this case, the quadratic form $\vec{v}\cdot\mathcal{A}^{-1}\vec{v}$, can be written as in Eq.\eqref{eq:fra:vtav}
\begin{equation}
        \vec{v}\cdot\mathcal{A}^{-1}\vec{v} =\beta^2S_2(\xi,\chi)+2\beta S_1(\xi,\chi)+S_0(\xi,\chi),
\end{equation}
where $S_2$, $S_1$ and $S_0$ depends now on both interface positions $\xi$ and $\chi$. We start by analyzing the behavior of $S_2$. Since  $S_2(0,0)\simeq N/\beta$, $S_2(0,N)\simeq N/\gamma$ and $S_2(N,N)\simeq N/\alpha$, we get for large values of $N$ the following expression
\begin{equation}
\label{piano}
    S_2(\xi,\chi)\simeq\frac{N}{\beta}+\left(\frac{N}{\alpha}-\frac{N}{\gamma}\right)\frac{\xi}{N}+\left(\frac{N}{\gamma}-\frac{N}{\beta}\right)\frac{\chi}{N}+C,
\end{equation}
where considerations similar to the previous ones confirm that the zero-th order term $C$ assumes the same value in Eq.(\ref{eq:fra:c0}).
Similarly, we can prove that for large values of $N$ the quantities $S_1$ and $S_0$ assume the same values obtained for the purely brittle fracture model, i.e.
\begin{align}
    S_1(\xi,\chi)&\sim \frac{2}{\beta+\sqrt{\beta^2+4\beta}},\\
    S_0(\xi,\chi)&\sim \frac{2}{2+\beta+\sqrt{\beta^2+4\beta}},
\end{align}
which are independent of $\xi$ and$\chi$.
Adopting these approximations in the expression for $q$, we obtain for $N\to\infty$
\begin{equation}
       q\sim\beta^2\left(\frac{\xi}{\gamma}-\frac{\xi}{\alpha}-\frac{\chi}{\gamma}+\frac{\chi}{\beta}\right)+\epsilon,
    \end{equation}
where $\epsilon$ is given by 
\begin{equation}
    \epsilon= \frac{\sqrt{\beta^2+4\beta}-\beta}{2}.
\end{equation}
This result corresponds to  Eq.(\ref{qsofttext}) of the main text. 
Finally, we can also find an approximated expression for $\det\mathcal{A}(\xi,\chi)$ that now depends on both $\xi$ and $\chi$.
Since  previous approximation for $\ln\det\mathcal{A}(\xi)$ was a linear function of $\xi$ linking the values obtained for the two homogeneous matrices at $\xi=0$ and  $\xi=N$, we can now assume that the approximation for $\ln\det\mathcal{A}(\xi,\chi)$ is a linear function in $\xi$ and $\chi$ passing through the three points identified by $(\xi,\chi) = (0,0),(N,0)$ and $(N,N)$.
Hence, we assume that
\begin{equation}
    \frac{\ln\det\mathcal{A}}{N} \sim a + b\xi + c\chi,
\end{equation}
where $a$, $b$ and $c$ are coefficients, which can be found with the assumptions
$a=\ln\tau_\beta$, $a+cN=\ln\tau_\gamma$, and $a+cN+bN=\ln\tau_\alpha$,
with
\begin{align}
    \tau_\alpha &= \frac{2+\alpha+\sqrt{\alpha^2+4\alpha}}{2},\\
    \tau_\beta  &= \frac{2+\beta+ \sqrt{\beta^2+4\beta}}{2},\\
    \tau_\gamma &= \frac{2+\gamma+\sqrt{\gamma^2+4\gamma}}{2}.
\end{align}
These values satisfy the relation $\tau_\beta<\tau_\gamma<\tau_\alpha$.
To conclude, we obtain the relation
\begin{equation}
    \frac{\ln\det\mathcal{A}}{N} \sim \ln\tau_\beta+\frac{\xi}{N}\ln\frac{\tau_\alpha}{\tau_\gamma}+\frac{\chi}{N}\ln\frac{\tau_\gamma}{\tau_\beta},
\end{equation}
which corresponds to Eq.(\ref{detsofttext}) of the main text.

\begin{acknowledgments}
The authors have been supported by Central Lille and Region Hauts-de-France under project MiBaMs. Moreover, the authors acknowledge stimulating scientific discussions with Giuseppe Florio and Giuseppe Puglisi.
\end{acknowledgments}
 

\end{document}